\documentclass[sn-basic]{sn-jnl}

\usepackage{tikz}
\usepackage{pgfplots}
\pgfplotsset{compat=1.18}
\usepackage{wrapfig}
\usepackage{graphicx}%
\usepackage{tabularx}
\usepackage{multirow}%
\usepackage{amsmath,amssymb,amsfonts}%
\usepackage{amsthm}%
\usepackage{mathrsfs}%
\usepackage[title]{appendix}%
\usepackage{xcolor}%
\usepackage{colortbl}%
\usepackage{textcomp}%
\usepackage{manyfoot}%
\usepackage{booktabs}%
\usepackage{algorithm}%
\usepackage{algorithmicx}%
\usepackage{algpseudocode}%
\usepackage{listings}%
\definecolor{shadecolor}{gray}{0.9}
\usepackage{framed}
\usepackage{breakurl}

\usepackage[colorinlistoftodos,prependcaption,textsize=small,disable]{todonotes}

\newcommand{\Section}[1]{Section~\ref{#1}}
\newcommand{\Figure}[1]{Fig.~\ref{#1}}
\newcommand{\Table}[1]{Table~\ref{#1}}

\newcommand{\RQ}[1]{\textbf{RQ\textsubscript{#1}}}

\begin{document}


\title{Code Review as Decision-Making - Building a Cognitive Model from the Questions Asked During Code Review}

\author*[1]{\fnm{Lo} \sur{Gullstrand Heander} \orcid{0000-0002-0695-4580}}\email{lo.heander@cs.lth.se}

\author[1]{\fnm{Emma} \sur{S\"oderberg} \orcid{0000-0001-7966-4560}}\email{emma.soderberg@cs.lth.se}

\author[1]{\fnm{Christofer} \sur{Rydenf\"alt} \orcid{0000-0003-1495-8263}}\email{christofer.rydenfalt@design.lth.se}

\affil[1]{\orgname{Lund University}, \orgaddress{\city{Lund}, \country{Sweden}}}

\abstract{
Code review is a well-established and valued practice in the software engineering community contributing to both code quality and interpersonal benefits. However, there are challenges in both tools and processes that give rise to misalignments and frustrations. Recent research seeks to address this by automating code review entirely, but we believe that this risks losing the majority of the interpersonal benefits such as knowledge transfer and shared ownership.

We believe that by better understanding the cognitive processes involved in code review, it would be possible to improve tool support, with out without AI, and make code review both more efficient, more enjoyable, while increasing or maintaining all of its benefits. In this paper, we conduct an ethnographic think-aloud study involving 10 participants and 34 code reviews. We build a cognitive model of code review bottom up through thematic, statistical, temporal, and sequential analysis of the transcribed material. Through the data, the similarities between the cognitive process in code review and decision-making processes, especially recognition-primed decision-making, become apparent.

The result is the Code Review as Decision-Making (\textbf{CRDM}) model that shows how the developers move through two phases during the code review; first an orientation phase to establish context and rationale and then an analytical phase to understand, assess, and plan the rest of the review. Throughout the process several decisions must be taken, on writing comments, finding more information, voting, running the code locally, verifying continuous integration results, etc.

Analysis software and process-coded data publicly available at:\\
\href{https://doi.org/10.5281/zenodo.15758266}{https://doi.org/10.5281/zenodo.15758266}
}



\maketitle

\section{Introduction}
\label{secIntroduction}

Code review is a well-established activity in modern software development valued for both quality assurance and interpersonal benefits~\citep{bacchelli_expectations_2013,sadowski_modern_2018}. On a global scale, with more than 28 million software developers~\citep{statista23} spending 10$\%$-20$\%$ of their working time reviewing code~\citep{sadowski_modern_2018,bosu2013impact}, more than 22-44 million hours are spent on code reviews daily. However, despite the importance of code review, there are significant misalignments between the tools used and the goals and actions of software developers, which reduces efficiency and adds frustration~\citep{soderberg_understanding_2022}. The code review process is also challenging for many teams with a range of common antipatterns~\citep{chouchen2021anti}. Improving code review tools and processes has huge potential benefits for the software engineering community. In addition to saving time on the code review itself, the 2023 DORA State of Devops industry report finds that teams with more efficient code review have up to 50\% higher software development throughput overall~\citep{dora2023}.

However, despite the potential benefits of improvements, the tools used for code review today are very similar to the first tools introduced in the early 1990s, for example, the ICICLE tool~\citep{brothers_icicle_1990}. Similarly to today's code review tools, such as GitHub\footnote{\url{https://github.com/}} or Gerrit\footnote{\url{https://www.gerritcodereview.com/}}, ICICLE was centered around a textual diff view where comments can be added by humans or by automated analysis. During the same time, we have seen huge developments in tools for writing, navigating, and understanding software~\citep{bird_taking_2023}. We should not leave code review behind. With the recent increase in AI assistance capabilities, there has been a growing interest in how to utilize AI-based assistance in software development tools~\citep{dora2024}. This trend also extends to AI-powered improvements in code review, specifically automated code review has received a lot of attention in recent years. For example,~\citet{Lu2023llama-reviewer} introduced LLaMA-Reviewer to automate the code review task. \citet{yu2024-carllm} presented Carllm for improved precision and clarity in automated code review. \citet{tang-etal-2024-codeagent} proposed CodeAgent, an approach in which multiple agents collaborate to find code quality issues. Google explored how to automate code review in their DIDACT project by training ML models on each of the sequential steps~\citep{froemmgen_resolving_2024}. There is also recent research on user experience improvements~\citep{heander2024codeblocks}, AI assistance frameworks~\citep{unterkalmsteiner2024help, heander_support_2025}, and innovative visualizations~\citep{krause-glau_visual_2024}. 

\begin{figure}
\fbox{\includegraphics[width=0.99\textwidth]{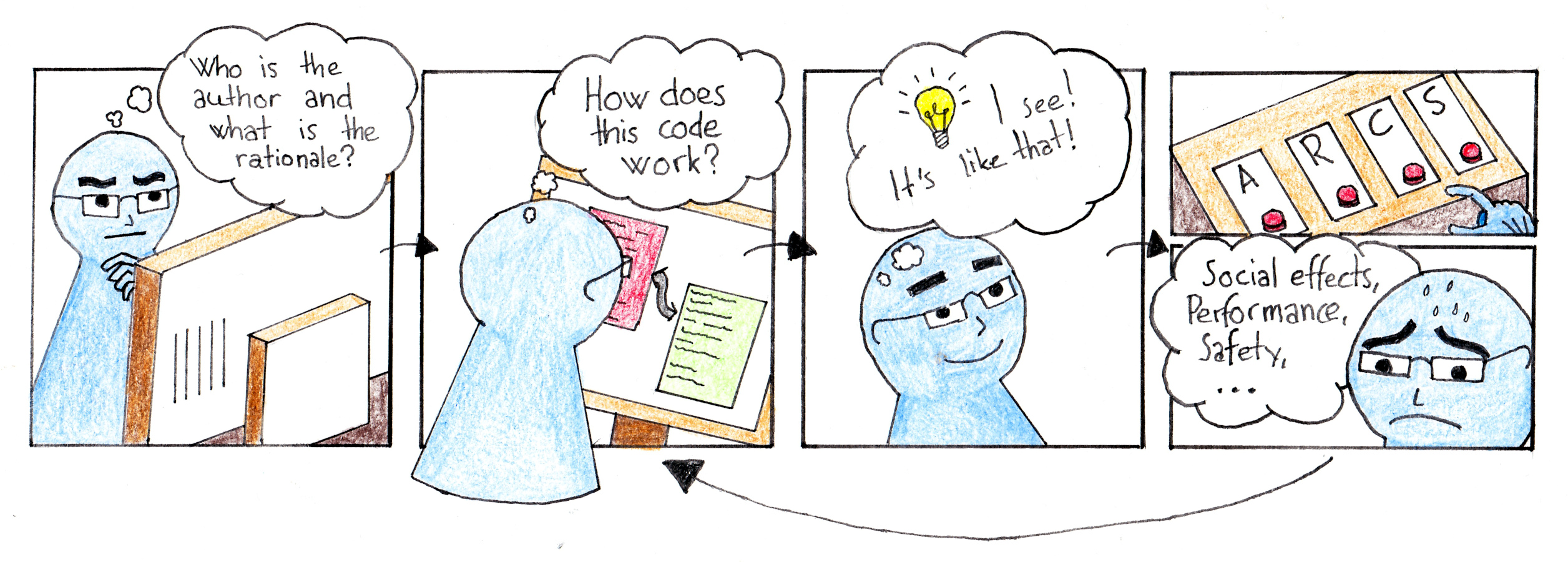}}
\label{figComic}
\caption{Illustration of steps and questions during code review. First, the developer orients themselves on the context of the code review, next asks how a part of the code change works, moves on to comprehending the code, and finally have to make a decision on how to proceed. Considering social effects, performance, safety, etc., should they \textbf{A}ccept, \textbf{R}eject, \textbf{C}omment, or \textbf{S}earch for more information. The reviewer iterates and looks at more code changes before reaching a final decision. \textbf{Takeaway:} Code review starts with an orientation phase, followed by an iterative comprehending-assessing-decision phase. Comprehending the code change is not the end goal, but rather a prerequisite for deciding how to handle the review.}
\end{figure}

Although the focus on automated code review is interesting, several of the benefits of code review, such as knowledge transfer and shared code ownership, are interpersonal and risk being lost if the activity is automated~\citep{heander_support_2025}. 
To improve code review processes and tools while preserving all the important benefits it is important to understand the activity and its challenges well. However, while the practical process of code review is well researched and described~\citep{sadowski_modern_2018,davila_systematic_2021}, there are few studies on the cognitive processes during code review. \citet{goncalves_code_2025} investigate how developers form and use their understanding of the changed code under review, building a model for code review comprehension~. This is interesting work contributing to the understanding of code review, but in this paper we want to look at a wider scope beyond comprehending the changed code and study the cognitive process of the code review activity as whole including choosing a review, writing comments, voting, looking for more information, etc. 

Building a theoretical model of cognitive processes in code review grounded in interviews and observations can facilitate improvements in code review tools and processes in several ways. By analyzing existing tools to see which parts of the cognitive process they facilitate or hinder, by using the model to reason about the effects of new tool ideas, or by adapting the code review process in your organization to better match the cognitive process of the developers. 

\subsection{Research Questions}

The goal of this study is to investigate the cognitive process during code review and how it can be modeled to increase our understanding of code review and guide future improvements to tools and processes. This leads us to the following main research question:

\begin{description}
    \item[\RQ{1}] \emph{How can the cognitive process of code review be modeled from a theoretical perspective?}
\end{description}

\subsubsection{Supporting Research Questions} 

Since the cognitive processes of the developers during code review are not directly observable, we must study their actions and behaviors and use that insight to theoretically model the cognitive process. A basic assumption of this work is that developers \emph{actions} when reviewing code are intentional, that is, that they are tied to some \emph{meaning} that gives them direction ~\citep{dourish2001action}, and that it is this \emph{intentional relationship} ~\citep{rosenberger2015field} that needs to be analyzed to understand the cognitive processes involved in code review. In line with the theory of planned behavior, we also assume that the actual behavior or actions of reviewers are predicted by their intentions ~\citep{ajzen1991theory}. Furthermore, we assume that \emph{questions asked during code review} gives an indication of the intention of the reviewer when the question is asked, and thus that they give valuable insights into the reviewers' intent or cognitive focus in different parts of the review process. 

To move from a state where you know nothing about the review to a state where you are ready to vote for accepting or rejecting the code change, many questions must be answered and different aspects of the code change understood. Similarly to Letovsky's study on questions during code comprehension~\citep{letovskyCognitiveProcessesProgram1987}, in this paper, we study the explicit and implicit questions asked during a code review in order to build a theory about the cognitive process of the reviewers. By analyzing the patterns of which questions are asked, when questions on certain themes occur and how they relate to each other, the observed data can build the foundation for a theory of the cognitive process during code review. This gives us the following supporting research questions:

\begin{description}
    \item[\RQ{2}] \emph{What questions do developers ask during code review?}
    \item[\RQ{3}] \emph{How do the questions asked during code review connect to each other and the overall code review process?}
\end{description}

To explore these research questions in a context as realistic as possible, we perform an ethnographic think-aloud study combined with interviews~\citep{Charmaz14, williamson_research_2006}. The study is conducted at the software tools department of a  multinational software company (\Section{sectionMethod}). In total, we observe 34 code reviews by 10 participants and analyze the results using thematic, statistic, and sequential analysis (\Section{secResults}). The analyses form a basis for the construction of a theoretical model of the cognitive process during code review, interpreted and illustrated in \Figure{figComic} (\Section{secTheory}).

\subsection{Contributions}

The contributions of this research are as follows.
\begin{itemize}
    \item A theoretical model of the cognitive process of code review closely relating code review to decision-making processes (\Section{secTheory}).
    \item A thematic and statistical analysis of the questions asked during code review (\Section{secResults}).
    \item Suggested directions for future work to apply the theoretical model to improve code review processes and tools (\Section{secDiscussion}).
\end{itemize}

\section{Background and Related Work}
\label{sec:background}

The method, interview quotes, analysis, and theory building in this study build upon an understanding of modern code review practices, challenges, social effects during code review, the terminology of Gerrit code review tools, as well as cognitive theories around decision-making processes.

\subsection{Code Review in Practice}

The term ``Modern code review'' was popularized by~\citet{bacchelli_expectations_2013}, where they defined it as \emph{``(1) informal (in contrast to Fagan-style), (2) tool-based, and that (3) occurs regularly''}. Over the years, the properties of modern code review have changed. Code review has become more formalized, and many teams have checklists and processes that, for example, define how to conduct the review, how many approvals are needed to merge the changed code and how quickly the review is expected to be done~\citep{goncalves_explicit_2022}. Modern code reviews are today even more centered around the code review tools used. The tools define much of the process, how the code is analyzed and read during the review, how comments and responses clarify or solve issues, and finally how the changed code is approved or rejected~\citep{sadowski_modern_2018,davila_systematic_2021}.

Industry and community practices also place an increasing emphasis on performing code reviews regularly and quickly. Since code reviews are mandatory in many teams and projects, high throughput of new features and bug fixes depends on code reviews being done as soon as possible. The industry report DORA Accelerate State of Devops finds that teams with faster code reviews have up to 50\% higher software delivery performance overall, marking it as an important area for improvements~\citep{dora2023}. As a reference, \citet{kudrjavets_mining_2022} analyzed code review times in eight different large open source projects and found a median time between submission and acceptance of less than 24 hours. \citet{sadowski_modern_2018} reports a median time of less than 4 hours between submission and acceptance, and a median of 4 code reviews per developer per week at Google.

Code review has been shown to have several benefits. Both for its nominal purpose of finding and reducing software defects, but also, importantly, for code improvement, finding alternative solutions, increasing knowledge transfer, building team awareness, improving the development process, sharing code ownership, avoiding build breaks, tracking rationale, and assessing teams~\citep{bacchelli_expectations_2013,sadowski_modern_2018}. Code review is also an efficient way to spread information, such as best practices or information about new features, in a software development organization. A recent study by \citet{dorner_upper_2025} shows that the information spreads to up to 85\% of the participating developers after an average of only 3 code reviews.

\begin{figure}[hb]
\centering
\includegraphics[width=0.84\textwidth]{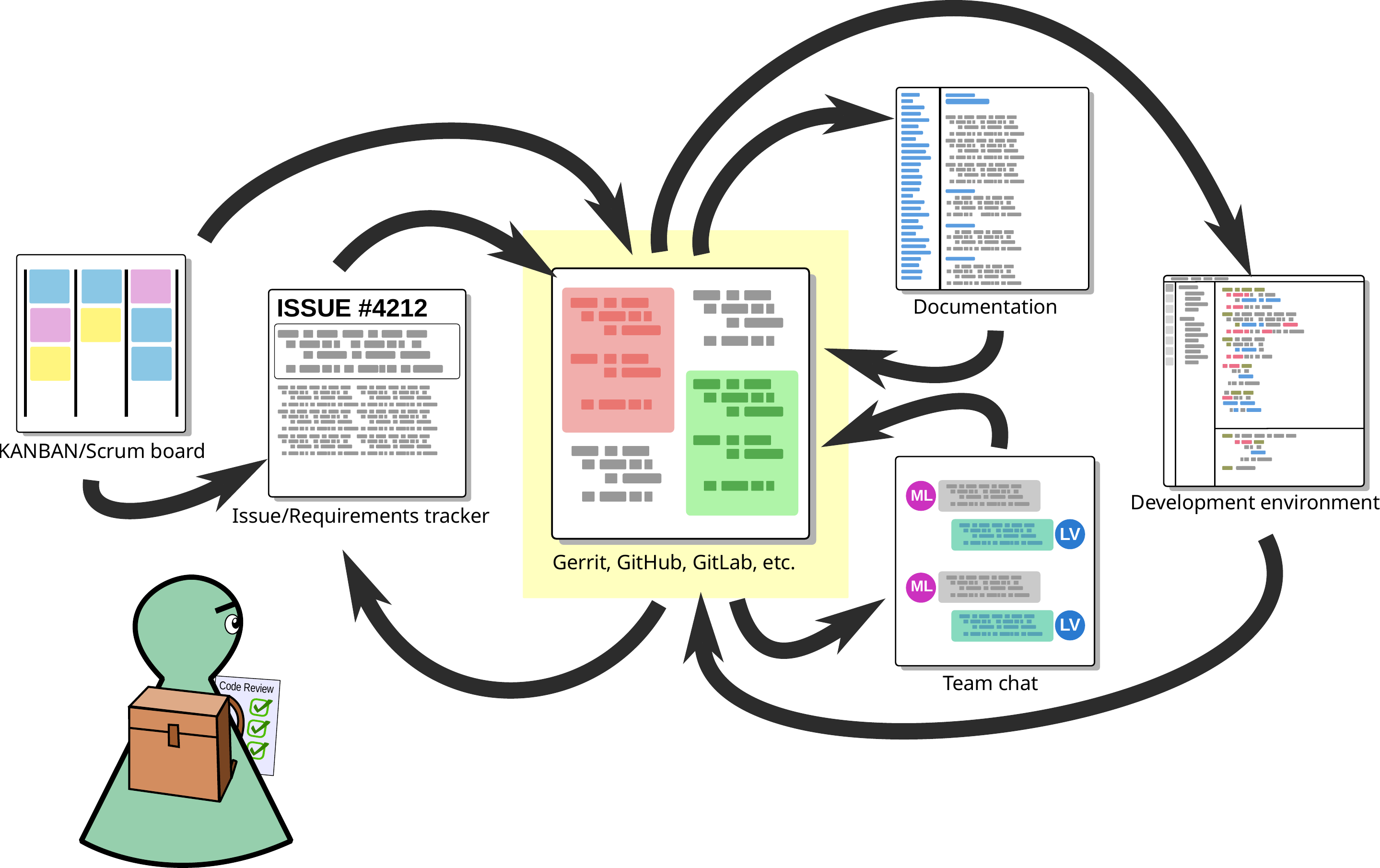}
\caption{Illustration of the experience of navigating between different tools during code review. The backpack and checklist represent the reviewer's experience and team processes respectively. \textbf{Takeaway:} The code review tool is in the center, but cross-referencing with several other tools is necessary to understand the full context of the code review.}
\label{figNavigate}
\end{figure}

There are also several challenges with modern code review practices and tools. Because of how current tools are designed, the reviewer needs to navigate between issue trackers, requirements databases, KANBAN boards, team chats, API documentation, continuous integration reports, etc., to gather the information needed to complete the code review. As illustrated in \Figure{figNavigate}, the reviewer must use their experience and the team processes to navigate between tools effectively and decide which steps are helpful and when~\citep{macleod_code_2018}. This experience takes time to build up, and becoming effective at code reviews in a new workplace can take up to a year~\citep{bosu_characteristics_2015}. Even with experience, it demands time, effort, and focus~\citep{soderberg_understanding_2022}. Other challenges include understanding the rationale for the code change~\citep{chouchen2021anti}, long response times, many repetitions of reviewing the same code change, and reviewing large code changes~\citep{dogan_towards_2022}.

\subsection{Social and Team Effects During Code Review}

There is recent work on improving code review practices and recommendations based on social and team effects in code review. \citet{pascarella_information_2018} investigate the information needs of the reviewers during the code review by analyzing discussion threads with questions and responses in open-source projects. They find seven main categories of information needs, such as rationale and code context, and recommend ways to improve code review by better meeting these needs.

\citet{lee2024understanding} studies anxiety and avoidance in relation to code reviews, both for code authors and reviewers. Their work outlines the main factors that contribute to code review anxiety and compel developers to avoid or procrastinate code review tasks, such as fear of judgment and criticism. They also present a CBT-based intervention that helps developers reduce anxiety after just a single session.

\citet{coelho_qualitative_2025} analyzes review comments and divides them into ``refactoring-inducing'' and ``non-refactoring-inducing''. They describe the factors leading to code refactorings in code changes and code review comments, since refactoring in the code review stage when the code is almost ready can be time-consuming. They give guidelines to researchers, practitioners, tool builders, and educators on how to better handle these situations and improve the code review process.

\subsection{Gerrit Code Review}
\label{secGerritCodeReview}

\begin{figure}[htb]
\includegraphics[width=0.9\textwidth]{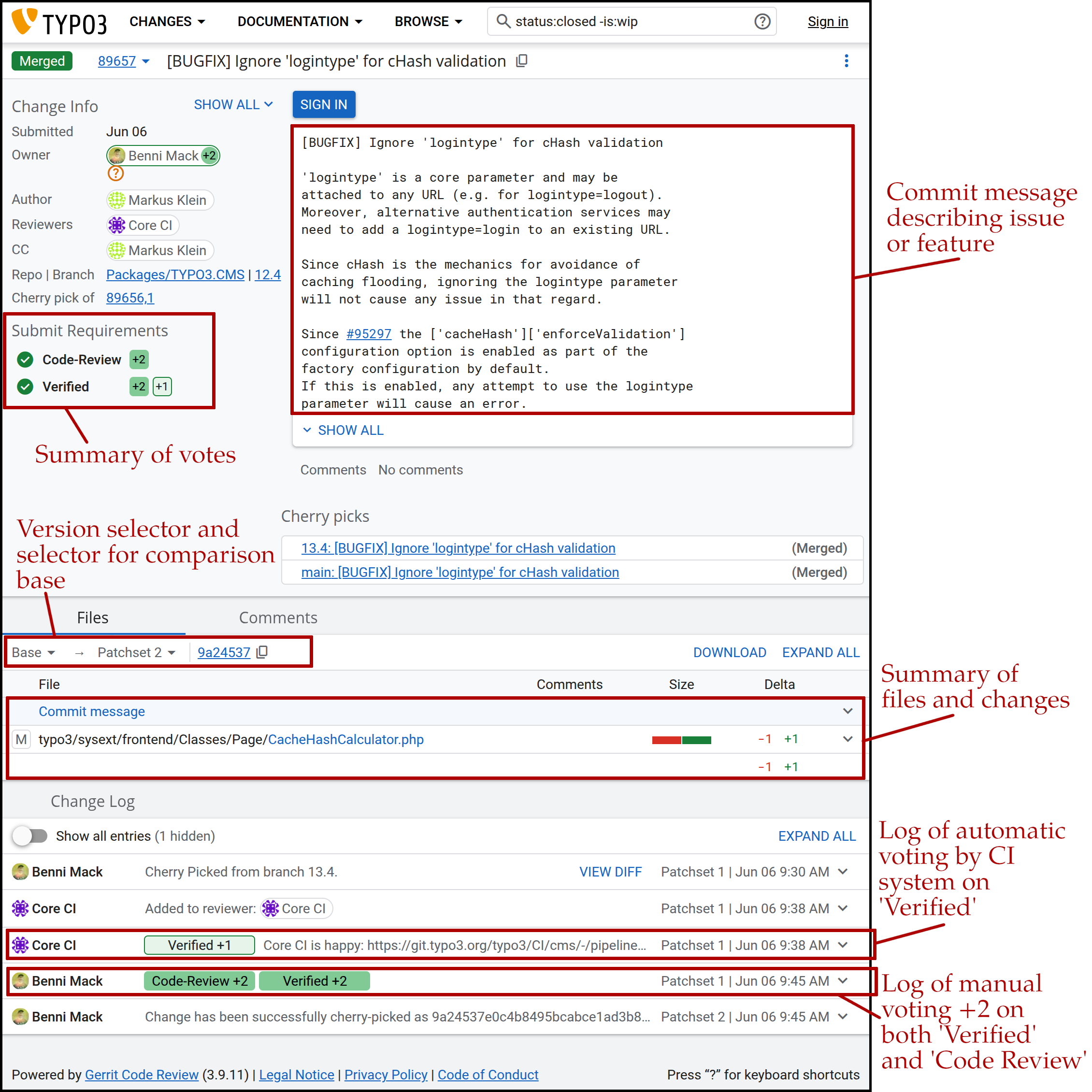}
\caption{Annotated screenshot of Gerrit user interface. \textbf{Takeaway:} Note how the `Change Info', `Submit Requirements', and `Commit message' sections gives an overview of the state and context of the code change. The log shows examples of both automated systems and human reviewers voting on different aspects.}
\label{figGerritUI}
\end{figure}

Gerrit Code Review\footnote{\url{https://www.gerritcodereview.com/}} is the open source code review platform used by the teams in this study. It is a widely used code review tool that has its origins in Google Mondrian, the code review platform used for many of Google's internal projects. When Google released the Android project as open source\footnote{\url{https://source.android.com/}}, they wanted a code review tool with features and workflow similar to Mondrian, but built with open source software and using Git\footnote{\url{https://git-scm.com/}} instead of Perforce for version control. \Figure{figGerritUI} shows an annotated screenshot of Gerrit's user interface. Some concepts and terminology from Gerrit show up in the quotes from the study.

\paragraph{Changeset and Patchset} A collection of one or more changed files with a commit message describing the rationale for the change is called a \emph{changeset} in Gerrit. A changeset can have many versions, called \emph{patchsets}, for example if the first version received some code review comments that led to updating the code and submitting a new version for re-review.

\paragraph{Voting} Gerrit is very flexible in how you can set up your workflows. While most code review software by default only allows the reviewer to accept or reject the changed code, Gerrit has a voting system instead. The administrator can create multiple labels to vote for and configure a range of numeric values for voting. By default the labels `Code Review' and `Verified' are available, signifying code review results and testing/verification results, respectively. Rules can be set up to allow merging the changed code only when certain voting scores are reached. In the setup used in this study, reviewers can make the decision to vote -2, -1, $\pm0$, +1, or +2 on the `Code Review' label, and each reviewer's vote accumulates to the total score on the code change. The code change must reach a total code review score of +2 or higher before it is possible to merge it into the main branch. This means that a vote of -2 will effectively block the code change from being merged, -1 will strongly discourage it, $\pm0$ will be a neutral vote, +1 means that you approve but you want someone more to take a look, and +2 signifies approval and ready to merge. Automated tests, linters, static analyzers, etc., can also vote on the code change but usually on labels like `Verified', `Formatted', etc., so their votes will not be confused with the scores from human reviewers. 

\paragraph{Commenting} Reviewers can decide to leave code review comments on individual lines of code as well as for the change as a whole, independently of how they choose to vote. Comments can be questions that need clarification, suggestions for improvements, pointing out potential issues, requests for fixes, or code for alternative solutions that the author can accept with just a click. When created, comments are marked as unresolved and must manually be marked as resolved by code change author or the reviewer who wrote the comment before the code change can be merged into the main branch. 

\subsection{The Recognition-Primed Decision Model}
\label{secRPD}

\begin{figure}[hbt]
\centering
\includegraphics[width=0.7\textwidth]{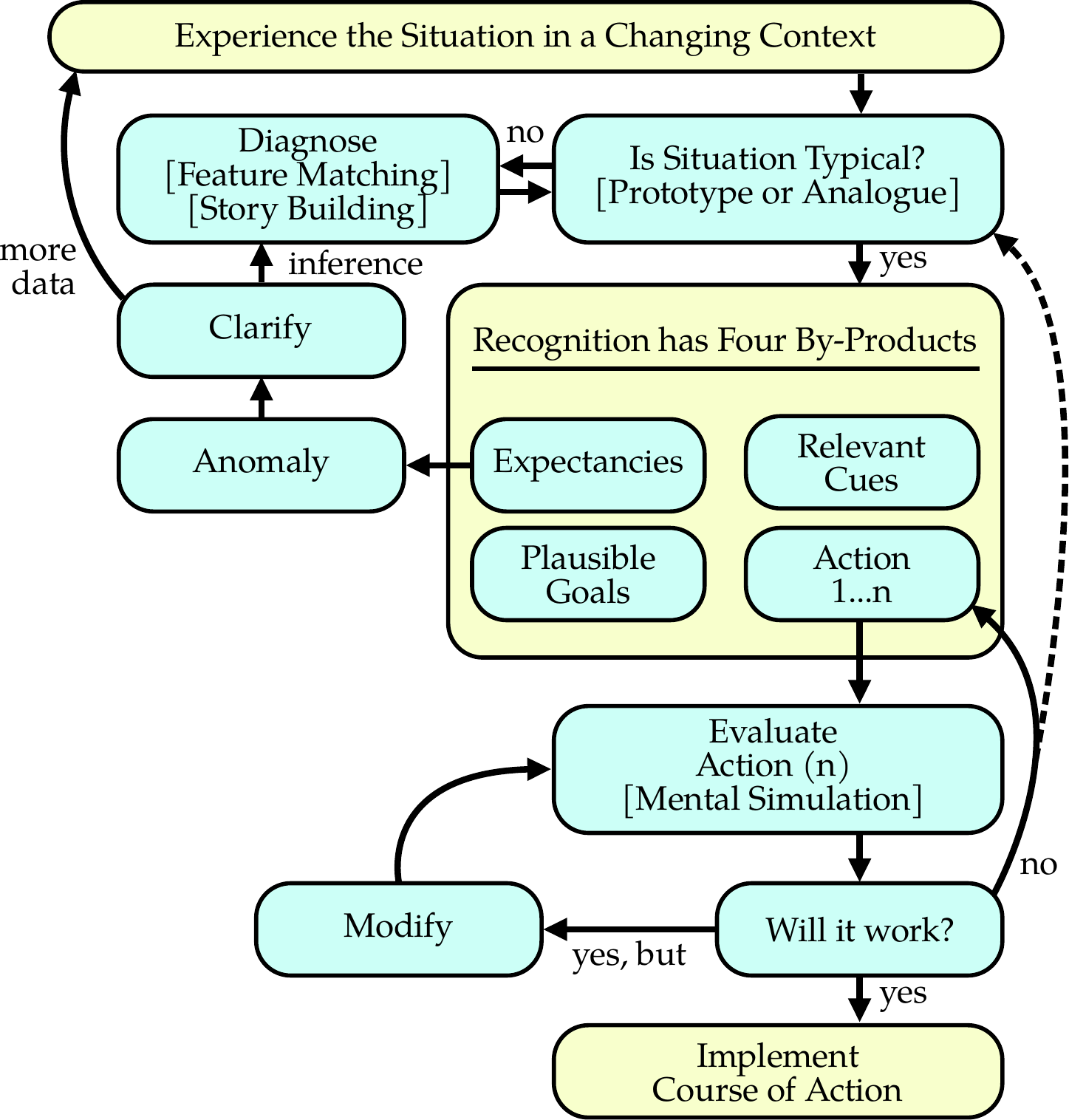}
\caption{Recognition-primed decision (RPD) model from~\citet{klein_sources_1998}. \textbf{Takeaway:} In contrast with previous models of rational decision-making, Klein models decision-making as a process that starts from experiencing the situation and recognizing similarities and differences to previous situations. From this recognition springs possible actions that are tested, first with mental simulation and then in practice.}
\label{figure:klein-rpd}
\end{figure}

Research on rational choice and decision-making has shown that in practice human rationality is quite far from living up to the standards of being absolute or globally rational (i.e., the ability to pick the objectively best choice). Instead, it should be considered bounded or local and dependent on the problem framing made by the decision maker and the currently available knowledge~\citep{simon1996sciences,simonRationalChoice1955,tversky1981framing,dekker2002reconstructing}.

However, even then, the decision-making process does not necessarily aim at the best possible solution. Rather, it has been shown that the decision maker applies a satisficing approach, that is, looks for the first solution available deemed "good enough" ~\citep{simon1996sciences}, and that heuristics and biases play an important part in the decision-making process ~\citep{tversky1974judgment}. 

Klein defines the ground-breaking and influential recognition-primed decision model (RPD model)~\citep{klein_sources_1998} from research on fireground and military commanders, who need to make critical decisions often and quickly. The RPD model differs from the more traditional rational choice strategy model~\citep{simonRationalChoice1955}, in that it does not list all available actions and their pros and cons. Instead, the RPD model describes how decision makers use their experience to, often subconsciously, identify analogous situations and take the first action that, by experience and mental simulation, seems likely to succeed. 

The RPD model (see \Figure{figure:klein-rpd}) is an iterative model that begins with experiencing a situation and evaluating if the situation is typical. If it is deemed typical in some aspect, i.e. recognized, this elicits expectancies, relevant cues, plausible goals, and typical actions. From here, there are two iterative flows possible. First, check if the perceived reality matches the expectations and if it does not go back, collect more data, and modify the story building until the expectations match reality. Second, evaluate possible actions by mental simulation and modify or discard the action until the first action likely to work is found. Then, this action is carried out, with the decision maker mentally prepared for some of the possible consequences. If the results differ from the mental predictions, the situation is re-evaluated and the decision maker carries out a different action they feel is likely to work under the new circumstances. The process repeats until the desired outcome is achieved, or there are no more actions likely to work.
\section{Methodology}
\label{sectionMethod}

\begin{figure}[h]
\centering
\includegraphics[width=0.94\textwidth]{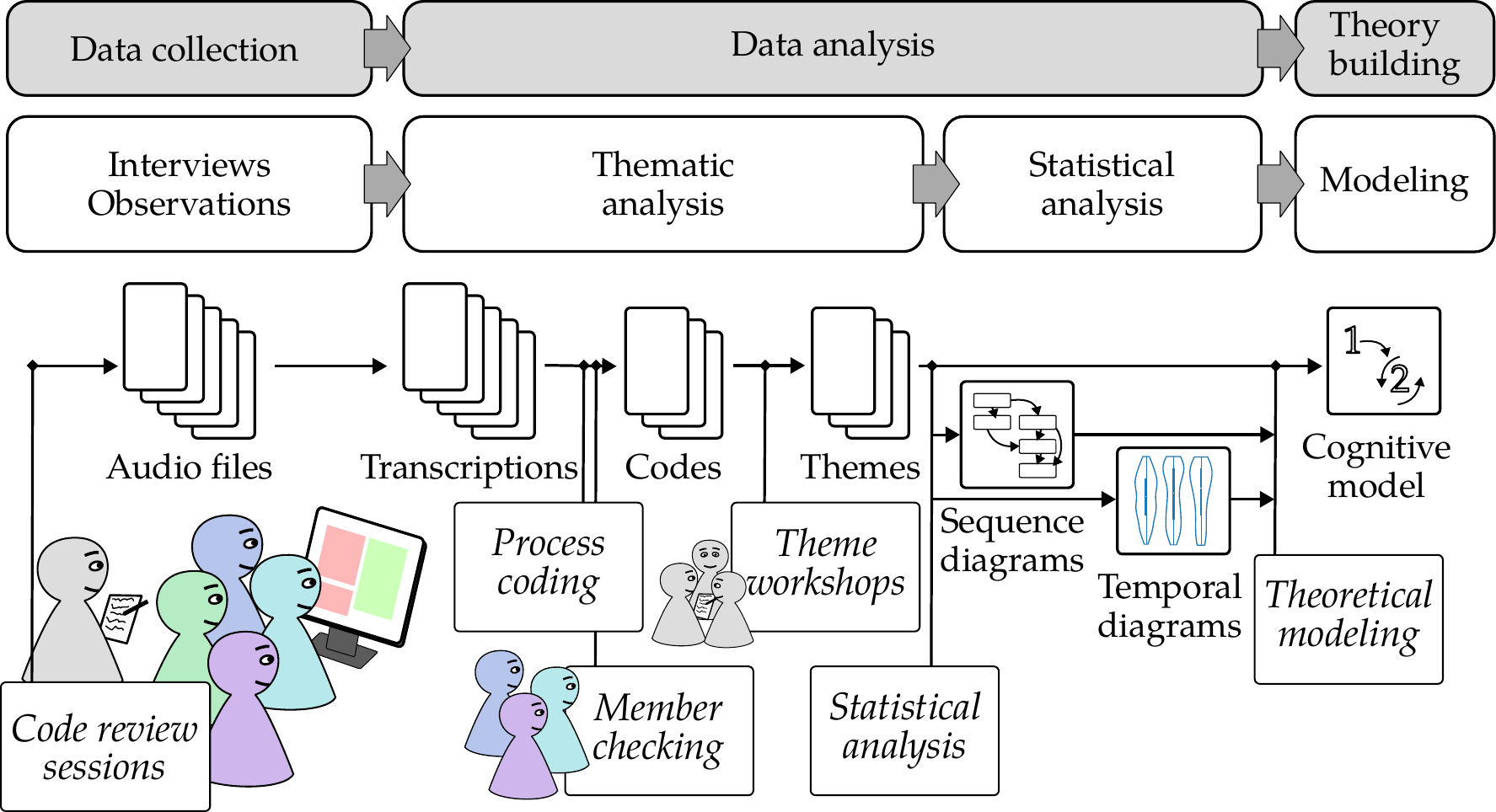}
\caption{Process and method overview. \textbf{Takeaway:} The theoretical output, a cognitive model of code review, is built upon data collection from real code review sessions followed by thematic and statistic analysis.}
\label{figProcess}
\end{figure}

In \Figure{figProcess}, we present an overview of the process and methods used in this study. The study design is based on constructivist epistemology applied to ethnographic methodology, as described by Williamson~\citep{williamson_research_2006}, with the purpose of co-constructing useful and applicable models and theories together with the study participants. As described by \citet{h_sharp_role_2016}, ethnography applied to software engineering is well suited to explore not only what practitioners do but also \emph{why} they do it. Sharp et al. also find that the results of ethnographic studies can deepen knowledge on social and human aspects of software engineering, inform improvements to software engineering tools, lead to process development, and point out directions for future research. All of which are goals for the contributions of this study. 

Specifically, to understand the questions asked during code reviews (\RQ{2}, \RQ{3}) and the corresponding cognitive processes (\RQ{1}), we designed an ethnographic think-aloud study combined with interviews~\citep{Charmaz14}. We worked with software developers in the industry and captured the participants' normal way of reviewing as closely as possible. Our goal was to help developers feel comfortable and view the study as an exploration, and not an evaluation, of their code review techniques, habits, and skills. To achieve this, the first author worked from the same office as the participants for several weeks to get to know the participants and to be available whenever someone needed to do a code review.

The first author sat right next to the software developers while they conducted real code reviews on their usual workstations, in their own workplace, and in their team context. Everyone was fully informed about the research, and we asked the participants to treat the first author as a newly hired developer and openly explain their ways of working. In this role, we could observe code reviews and the questions asked by developers and document their thought processes and strategies to choose code reviews, ask and answer questions, write review comments, and conclude the code reviews. 

The first author has worked as a software developer, project manager, team lead, etc., in similar companies for over 15 years and has extensive first-hand experience with code reviews. This background contributed to creating good rapport between the researcher and the participants. Encouraging participants to be more detailed, vocal, and open in describing their code reviews, backgrounds, and processes.

\subsection{Study Context}

To achieve depth in our interviews and code review sessions, we presented the study idea to a multinational software company where we already had an active industry-academia collaboration agreement. The second author contacted the company to ask for a meeting in which we could present the study, and after the company accepted the invitation, the first author gave a presentation about the background of the study, its goals, and its methodology.

In agreement with the company, we decided to focus the study on their tools department. The tools department has around 30 developers with different levels of experience divided into 8 teams and working in several programming languages. This department was chosen because it would give a broad view of different practices and levels of experience and because it works almost exclusively with open-source software, allowing us to conduct the study and report the findings more openly.

\subsubsection{Code Review Process}

All teams use Gerrit as their code review platform (\Section{secGerritCodeReview}), and code reviews are mandatory for all code changes. Depending on the size of the team, one or two other team members must approve every code change before it can be merged. The majority of the repositories have automatic linting, code formatting, unit tests, and continuous integration (CI), eliminating the most basic code review comments and allowing reviewers to focus on higher-level issues. The process requires all team members to review code at least once a week and to try and check their code review inbox daily so that code changes do not get stuck for too long waiting for review.

\subsection{Participants}
\label{sectionMethodParticipants}

\begin{table}[!htp]
\caption{Role, experience, and weekly time spent on code review for the participants in the study.}
\label{tableDemographics}
\footnotesize
\centering
\begin{tabular}{l|l|r|r|r}
\toprule
\textbf{Participant} & \textbf{Role} & \textbf{Code review exp.} & \textbf{Role exp.} & \textbf{Weekly code reviews} \\
\midrule
P01&Developer&3 years&3 years&1 hour\\
P02&SW Architect&14 years&9 years&2 hours\\
P03&Developer&14 years&9 years&2 hours\\
P04&Developer&1 year&1 year&0.75 hours\\
P05&Developer&6 years&1 year&5 hours\\
P06&Developer&2 years&2 years&5-10 hours\\
P07&Team Lead&5 years&5 years&2-3 hours\\
P08&Team Lead&16 years&7 years&6 hours\\
P09&Developer&11 years&11 years&4-8 hours\\
P10&Developer&17 years&13 years&5 hours\\
\midrule
\textbf{Average}&&8 years&6 years&4 hours\\
\textbf{Std.dev.}&&6 years&4 years&2 hours\\
\bottomrule
\end{tabular}
\normalsize
\end{table}

During the duration of the study, 10 software developers, from the tools department mentioned above, participated in interviews and code review sessions. Their role, experience in the role, experience with code reviews, and average time spent on code reviews weekly are found in \Table{tableDemographics}. No participants left the study during or after field work. The participants are all of Swedish nationality and have at least a Bachelor degree. The teams involved each have 2-7 developers and work according to the agile software development methodology. In accordance with the team process, all participants did code reviews at least every week with many of the participants reviewing code every workday.

\subsection{Data Collection}
\label{method:data_collection}

To test our study design and the interview protocol, we conducted a pilot interview and code review session at a small local software company. The pilot session went smoothly and gave us no reason to change the study setup. Data from the pilot study were kept for reference and comparison, but excluded from data analysis and results.

For each participant, we collected informed consent for the participation in the research study and interviewed them about their background, experience with code review, and role in the team. Whenever a participant had a pending code review to carry out, the first author sat next to them, observing their work and asking them to think aloud about what they were doing during the code review sessions. We asked questions or noted their actions aloud to record them in the sound file and to encourage the participants to explain and reflect on what they were doing and why. The code review sessions and the interviews were recorded using a dictaphone. 

\subsection{Data Analysis}
\label{methodDataAnalysis}

To analyze the material in this study, we used the principles of Williamson's constructivist ethnographic research~\citep{williamson_research_2006} and thematic analysis following the general process described by Braun and Clarke~\citep{braun_using_2006, clarkeThematicAnalysis2017}:

\begin{enumerate}
    \item \emph{Familiarizing yourself with your data:} We transcribed and annotated the recorded material and performed member-checking to validate the data.
    \item \emph{Generating initial codes:} We coded the transcriptions using process coding.
    \item \emph{Searching for themes:} Two authors organized the process codes into themes independently.
    \item \emph{Reviewing themes:} All authors reviewed and analyzed the themes until interpretative convergence.
    \item \emph{Defining and naming themes:} We refined the theme naming by studying excerpts from transcriptions.
    \item \emph{Producing the report:} We analyzed themes using statistical and sequential analysis.
\end{enumerate}

\subsubsection{Data Transcription}

The first author manually transcribed the sound files from all interviews and code review sessions into Markdown-formatted text. The text files were annotated with information about the participant, the time and date, and the beginning and end of each code review session. The interviews were conducted in Swedish mixed with many software engineering terms and anglicisms. Care had to be taken when transcribing the material to preserve the meaning and intent faithfully. Due to data privacy agreements with the company and for language reasons, we did not use automated transcription software. 

\subsubsection{Participant Feedback \& Member-Checking}
\label{methodMemberChecking}
To verify emerging themes from the data and establish an approach for coding the finished transcripts, we invited all participants to a member-checking focus group meeting~\citep{birt_member_2016}. The meeting was planned and facilitated by the first author and participants P01, P02, P05, P06, P09, and P10 attended. The meeting was recorded with a tabletop dictaphone to ensure a clear recording of everyone's voice.

During the meeting, we presented examples from the transcriptions for each emerging theme. The participants then discussed whether they recognized the situations in the quotes and what their experiences were like in similar situations. There were also general discussions on the processes and challenges in code review. The first author took notes that were reviewed during the meeting by the participants. These notes and the audio recording of the meeting were combined into meeting minutes that described the points to be taken into account during the continued analysis of the transcribed data.

\subsubsection{Initial Coding}
\label{method:coding}

\begin{figure}[hbt]
\centering
\includegraphics[width=0.99\textwidth]{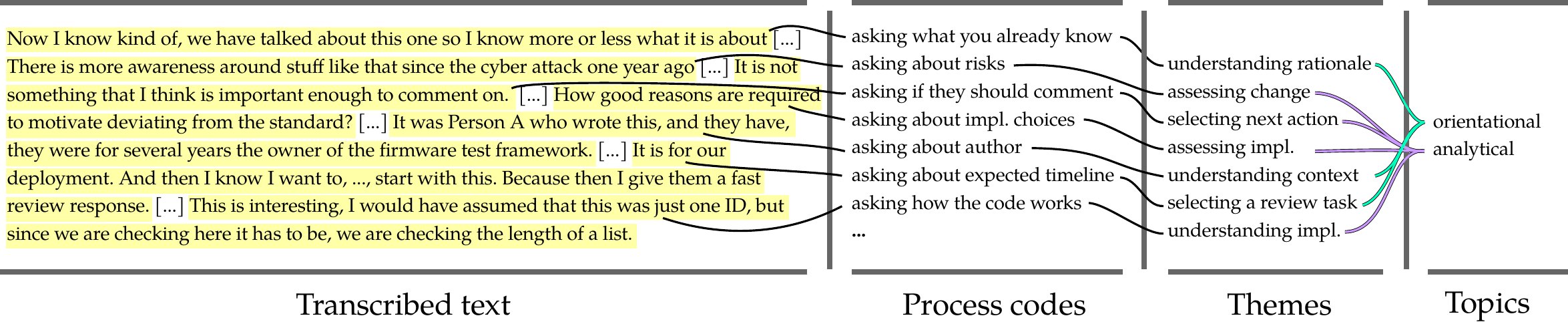}
\caption{Example of process coding showing excerpts of transcribed text, a small selection of process codes, and mappings onto themes and topics. \textbf{Takeaway:} The transcribed text is abstracted into process codes, process codes are abstracted into themes, and finally themes are abstracted into topics.}
\label{figCoding}
\end{figure}

Coding, in qualitative methods, is the practice of assigning a label or code to each sentence or segment of a text~\citep{Charmaz14}. The purpose is to increase the abstraction level of the text to facilitate the data analysis and comparison of different parts of the material. Different ways of coding will highlight different aspects of the original data. Coding is often done repeatedly, first by abstracting text into codes, then systemizing the codes into themes, and then grouping the themes into topics (see \Figure{figCoding}). Since the study aims to understand the developers' cognitive process during code review, we chose process coding both for the initial coding of the transcribed data and for constructing the themes. The first author did the initial coding, with authors two and three reviewing the coding and suggesting changes.

Process coding is a methodology where the first word in every code must start with a verb in the gerund form (ending with \emph{-ing} in English)~\citep{saldana_2015}. This form of coding is designed to capture the action and intention behind each coded segment~\citep{Charmaz14}, making it especially suited to uncover underlying questions and processes. Its name comes from how it results in a timeline of actions, a process.

\subsubsection{Identifying Themes}

The first and second authors independently grouped the process codes into themes to get two contrasting starting points. Based on these two sets of themes, all three authors discussed and worked through differences in workshop meetings. In these meetings, process codes, suggested themes, and excerpts from the transcripts were used. We drew connections between the sets of themes, regrouped process codes, and discussed until we reached interpretive convergence~\citep{saldana_2015}. Finally, using the transcripts, we refined the naming of the themes to capture and communicate the intent of the segments they covered.

\subsubsection{Statistical Analysis} 
\label{method:data_analysis}

When codes and themes were established, we wrote a Python program to analyze the data using temporal and sequential analysis. The sequential analysis determined the transition rate from each theme of questions to another. We gathered the transition rates into a transition table and constructed a sequence diagram of the code review process using the top 3 most common incoming and outgoing transitions with a rate higher than 0, see \Figure{figure:rel_network}.

For temporal analysis, we positioned each theme in relation to the start and endpoints of each code review. Sometimes, code reviews were paused or interrupted due to meetings, the end of the work day, or other more urgent code reviews. In the analysis program, we detected these interruptions and gathered the codes and themes for each code review in a linear flow. We normalized the duration of each review, which ranged from a couple of minutes to almost an hour, to a scale from $0$ at the beginning of the review to $1$ at the end. We analyzed the distribution of each theme, as well as the process codes contained, over the duration of the code reviews.

\subsection{Ethical Considerations}

Contributing to the research study would take significant time and effort for the participants, and neither they nor the company could be reimbursed or compensated for this. Participation was done with informed consent and on an explicitly voluntary basis. For participants, the upside would be learning something new about how they perform code reviews by explaining it to someone else, a change of pace in their workday, and the feeling of contributing to research.

During sessions and interviews, statements might come up that could be negatively interpreted by colleagues or the employer. To protect the participants, we pseudonymized all interviews and keep audio recordings and transcriptions confidential except for excerpts used to exemplify codes and themes. To allow other researchers to verify and replicate our study, it would be ideal to publish all collected data, but as discussed above, this is not possible for privacy and confidentiality reasons. The interview protocols, the data analysis program and the process-coded data set are available in the replication package (\Section{matterReplicationPackage}). 
\section{Results}
\label{secResults}

\begin{table}[hbt]
\caption{Overview of code review sessions from which we collected data. Note that a coded segment roughly corresponds to a paragraph in the transcription.}
\label{tableSessions}
\footnotesize
\centering
\begin{tabular}{l|r|r|l}
\toprule
\textbf{Participant} & \textbf{\# Segments} & \textbf{Duration (min.)} & \textbf{Outcome}  \\
\midrule
P01&27&26&vote $\pm0$ with comments\\
P01&7&4&vote +2\\
P01&17&20&vote $\pm0$ with comments\\
P01&4&2&vote +2\\
P01&2&1&vote +2\\
\midrule
P02&98&48&vote +1 with comments\\
P02&7&9&vote +1 with comments\\
P02&32&6&vote +1\\
P02&166&75&vote -1 with comments\\
\midrule
P03&141&49&vote -1 with comments\\
P03&15&5&vote +2\\
P03&7&2&vote +2 with comments\\
\midrule
P04&45&42&vote +1 with comments\\
\midrule
P05&50&38&vote +1 with comments\\
P05&15&21&vote +1\\
P05&15&6&vote +2\\
P05&9&6&vote +2\\
P05&49&32&vote +1 with comments\\
P05&15&37&vote +1 with comments\\
\midrule
P06&70&49&vote $\pm0$ with comments\\
\midrule
P07&75&41&vote $\pm0$ with comments\\
\midrule
P08&20&6&vote $+2$ with comments\\
P08&34&9&vote +2\\
P08&37&17&vote +1 with comments\\
\midrule
P09&14&9&vote +1 with comments\\
P09&25&17&vote +2 with comments\\
P09&13&7&vote -1 with comments\\
\midrule
P10&16&4&vote +1 with comments\\
P10&16&7&vote +1\\
P10&43&26&vote -1 with comments\\
P10&3&1&vote +1\\
P10&61&21&vote $\pm0$ with comments\\
P10&4&1&vote +2\\
P10&7&4&vote +1\\
\midrule
\textbf{Total}&1159&648&\\
\textbf{Average}&34&19&\\
\textbf{Std.dev.}&38&18&\\
\bottomrule
\end{tabular}
\normalsize
\end{table}

The study includes 10 participants (\Table{tableDemographics}); 7 out of 10 described their role as `Developer', 2 as `Team Lead', and 1 as `Software Architect'. The minimum work experience among the participants, both in their role and in code reviews, was 1 year. The most experienced participant had worked with code reviews for 17 years and in their current role for 13 years. The average was 8 years of code review experience and 6 years of experience in their current role.

We gathered data from a total of 34 code reviews (\Table{tableSessions}). The mean code review duration was 19 minutes with a standard deviation of 18 minutes. The shortest review took 1 minute, while the longest one took 75 minutes. During all the recorded code reviews, we process-coded a total of 1159 segments, meaning an average of 116 coded segments per participant and an average of 34 segments per review. 

The recorded sessions show a clear majority of positive votes (\Table{tableDemographics}); 26 of 34 code reviews receive a vote of at least +1. Often positive votes are given even when the reviewer wrote code review comments that they wanted to be addressed, either by updating the code or explaining the current implementation. Many participants say that the teams have a culture of trust and in general vote for merging the code and trust the author to address comments in a good way without needing re-review.

\begin{quote}
    \emph{``To force them to fix this little issue and then them having to wait for me to get back and approve feels very silly. So I leave a comment, set +2. [\ldots] So if, because I trust them to fix it, I don't have to come back and look at it [again].''} \textemdash P06, post code review interview
\end{quote}

\begin{quote}
    \emph{``You can trust that people will fix it in a satisfactory way. I don't need to look at it again. It is just a waste of time. Especially if you have several people, it becomes, like, it adds lead time. So then you can vote +1 or +2 if you anyway think that `oh, I trust that this person will do, do something good regarding my comment'. And then we have configured it so [\ldots] you cannot, you cannot submit the change when you have unresolved comments.''} \textemdash P02
\end{quote}

\begin{figure}[!htb]
\centering
\includegraphics[width=0.99\textwidth]{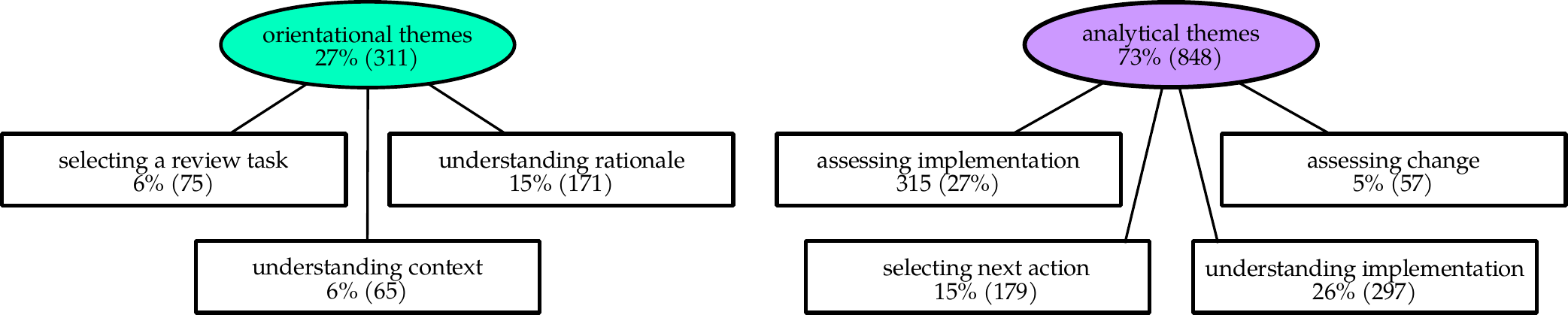}
\caption{Topics, their contained themes, and the number of occurrences in the data set. \textbf{Takeaway:} Over 70\% of the questions fall under the analytical themes indicating a majority of the code review effort is spent here.}
\label{fig:theme_map}
\end{figure}

In our thematic analysis (see \Section{methodDataAnalysis}) we constructed 157 process-codes to abstract explicit or implicit question underlying the 1159 segments. During the process-coding, we reached data saturation after coding around 3/4 of the transcribed material and after this very rarely encountered new process codes. We grouped the process-codes into 7 themes encoding commonality in underlying intentions. The detailed mapping from process codes to themes can be found in the replication package (\Section{matterReplicationPackage}). The three most common themes are \emph{assessing implementation}, \emph{understanding implementation}, and \emph{selecting next action}, which were found 315, 297, and 179 times, respectively, in the transcriptions. 

In turn, the themes are grouped into two topics: \emph{orientational themes}, and \emph{analytical themes}. \Figure{fig:theme_map} shows the topics, the enclosed themes, and the number of occurrences in the process-coded segments. The orientational themes revolve around how to make sense of the context, framing, and rationale of the review, while the analytical themes are about understanding the changed code, evaluating it, planning the next action, and making a decision about the change as a whole. The analytical themes are the most common and make up 73\% of the process-coded segments, with the orientational themes covering the remaining 27\%. 

\subsection{Orientational Themes}

In the transcribed data, we found many different kinds of questions asked to explore the context, framing, and rationale of the code review; orienting the code review task in relation to author, repository, expected timeline, programming language, rationale for the code change, available time for the reviewer, etc. This topic spans three themes: \emph{selecting a review task}, \emph{understanding context}, and \emph{understanding rationale}.

\subsubsection{Theme: Selecting a Review Task}

Questions about how to pick which code change to review. This involves asking about the amount of time you have available as a reviewer, the urgency of different code changes, social factors, competence, and interests. Many reviews begin from the list showing all code changes that are waiting for code review, where the reviewer picks one based on available time, code change size, and urgency.

\begin{quote}
\emph{``I get an overview. And here I can somehow, now it is only 3 reviews, but for the case where it would have been, I don't know, say 10 reviews. And somewhere, I have, you have 8 hours per day. Then I have to prioritize somehow what, what it is I will look at. [\ldots] I try to somewhere check a bit what it is that\ldots yes, but what is kind of the most important? Everybody wants to get their code out, but what, what is the prio-order? If I, like, know Person C is working on something that is, that they have been doing for a long time and they really want to get it out. And it is the last thing in the stack, then I will rather take that than for example this. Then I know the other one is not, is like, it is not as urgent.''} \textemdash P05
\end{quote}

\begin{quote}
\emph{``And then if there are several things [to code review] you can maybe make time for a couple of the small ones but have to leave the big ones, or you take one big [code review] and leave the small ones.''} \textemdash P07
\end{quote}

\noindent A more experienced participant with a software architect role also factored in if their expertise was required, or if someone else could review the code instead.

\begin{quote}
\emph{``So the choice of what I will look at is kind of the combination of, like, urgency and if I, if it is like, so to speak, are they waiting for me or are there others that can do it?''} \textemdash P02
\end{quote}

\noindent One participant kept it simple and usually picked the code change that had been waiting for the longest time.

\begin{quote}
\emph{``I usually take the oldest.''} \textemdash P01
\end{quote}

\subsubsection{Theme: Understanding Context}

Questions to orient the change and the review task in regards to who the author of the change is, what programming language it is written in, which repository it is, how long ago the change was posted, the expected timeline for deploying the change, previous reviews, related work, etc. In one review, the reviewer asks about who the author is and their background and concludes that the changed code will probably have tests in place:

\begin{quote}
\emph{``It was Person A who wrote this, and they have, they were for several years the owner of the firmware test framework. [\ldots] So that's good, you don't have to nag them to write tests at least.''} \textemdash P02
\end{quote}

\noindent During another session, a reviewer asks about repository, author, and author's recent work.

\begin{quote}
\emph{``I check repo and person. Because then I know a little, I know for this repo it is mostly Person B working in it right now. And then I can, like, infer that, ok, I know what Person B is up to and that gives me some context still\ldots''} \textemdash P05
\end{quote}

\noindent One reviewer checks for previous code review comments.

\begin{quote}
\emph{``So that, also the others have had some, a number of comments already.''} \textemdash P09
\end{quote}

\noindent For a code change with previous reviews and multiple versions (patchsets in Gerrit terminology), one participant asks which version they have seen before and sets that version as the base version for comparisons.

\begin{quote}
\emph{``I would probably do the same thing here\ldots no, right, this is a re-review so then I think I will compare with patchset 1 that is the, the one I reviewed last time.''} \textemdash P02
\end{quote}

\subsubsection{Theme: Understanding Rationale}

Understanding \emph{why} a change has been made and what goals it is trying to achieve. In one example, the reviewer directly asks about the rationale and how it compares to their expectations.

\begin{quote}
    \emph{``\ldots otherwise it would have been interesting to know `what are we trying to solve here, really?'. And it seems like maybe the scope has become a bit bigger, now they have done a bunch of other stuff''} \textemdash P03
\end{quote}

\noindent For another participant, it is the first thing they ask about when starting a code review.

\begin{quote}
\emph{``So then you will, first of all, check if I understand this. Because that is often a thing for me with his stuff. I mean, do I know what he wants to do here?''} \textemdash P04
\end{quote}

\noindent Many participants investigate the rationale by reading the original issue in the issue tracker (JIRA in the case of this company).

\begin{quote}
\emph{``Right, let's see what it says\ldots [reading quietly, from JIRA] Yes, so then I see that this is also just a part towards making us more automated.''} \textemdash P10
\end{quote}

\begin{quote}
\emph{``If we have an issue then it is, then I often think it is good to go in here and check because you can get some background. A bit more. People are different in how much you want to write in your commit-message and stuff.''} \textemdash P04
\end{quote}

\begin{quote}
\emph{``OK, now we go to the JIRA-issue.''} \textemdash P01
\end{quote}

\subsection{Analytical Themes}

Many questions in the transcripts involve analytical themes aiming towards solving the code review task, i.e., finding defects, writing comments, voting, etc. These are questions that seek to understand and evaluate the implementation details of the changed code, plan the next action to take in the code review, and also to evaluate the change as a whole. This topic spans four themes: \emph{assessing change}, \emph{assessing implementation}, \emph{selecting next action}, and \emph{understanding implementation}.

\subsubsection{Theme: Assessing Change}

In this theme, there are questions about whether the change as a whole, regardless of the specifics of the implementation, meets the reviewer's expectations. The reviewer questions security, performance, interoperability, rationale, compatibility with future development plans, etc. One reviewer asks about the impact of the changed code: 

\begin{quote}
\emph{``Because this is that kind of change. Here, it is the total opposite of the last one, this is something that affects a lot of people. It is a critical piece (of code).''} \textemdash P10
\end{quote}

\noindent One example involves questioning what would happen if the change was deployed as-is.

\begin{quote}
\emph{``In practice we could have rolled out this change today. It is just that we probably would have gotten support tickets asking `how does this actually work?' ''} \textemdash P07
\end{quote}

A common example is general reasoning about the risks involved with the change as a whole, regardless of the specific implementation choices.

\begin{quote}
\emph{``The thing is, what are the risks with this change? Either it doesn't work for unknown reasons or it could, it would maybe create, maybe create a huge amount of tickets in JIRA. It is maybe not great to overload JIRA either. But we would have noticed it pretty quickly\ldots''} \textemdash P04
\end{quote}

\subsubsection{Theme: Assessing Implementation}

Asking questions about whether the implementation follows the code conventions for the project, is readable, is correct, has bugs, meets the rationale, etc. An example is a session where the reviewer dislikes the implementation choices in the tests. 

\begin{quote}
\emph{``I'm not very fond of what he has done here, in that he uses his own struct-type for the tests.''} \textemdash P02
\end{quote}

\noindent Another reviewer questions the safety of the import statements used.

\begin{quote}
\emph{``So now we have imported something from the backend, from a backend project, so to speak. It is maybe not always safe to do that from the frontend without, what you call like, there is something called isomorphic javascript. [..] It is doubtful if you can do what is done here.''} \textemdash P03
\end{quote}

\noindent Asking about or commenting on error handling and system messages is also common.

\begin{quote}
\emph{``To just say `failed to run command' is not a very good error message in my opinion.''} \textemdash P02
\end{quote}

\noindent Questions verifying that the code looks reasonable and follows expectations is also frequently found.

\begin{quote}
\emph{``Here we check, check the format, check if things look reasonable. `New host replace old machine', ok. And this looks relatively reasonable.''} \textemdash P05
\end{quote}

\subsubsection{Theme: Selecting Next Action}

Questions about what the next step in a review is. For example, on whether to read related documentation, run the code locally, read the code again, talk to the author, reference the issue tracker, and more. One strategy encountered in the code review sessions is to look for the entry point and read the code in the order it is called during execution.

\begin{quote}
\emph{``Then I try to identify, if I look at the file list, to find, like, the top of the call stack. So you do not start by going deep down into, like, a leaf function.''} \textemdash P02
\end{quote}

\noindent Asking if they should write a comment or not is common and often involves reasoning about the implementation.

\begin{quote}
\emph{``Do we want a database connection directly from the service layer? It might be that we have that in other places but that I have just forgotten. In that case, like, it is like that. But we will write a comment on it.''} \textemdash P06
\end{quote}

\noindent Choosing how to vote involves different strategies. One participant checks the overview over all code review comments to inform their decision and to remember questions they had about the code change.

\begin{quote}
\emph{``And now, sometimes I have not really decided how I will, what I will do [when voting]. But then you can get an overview here over what, where you can see, but is that\ldots? [looking at a preview of all code review comments] Mostly, mostly a lot of small\ldots and then it was, what was the thing again\ldots? Something I did not like a lot that I was going to look at later, what was it now again\ldots?''} \textemdash P02
\end{quote}

\noindent Reviewers often take into account the size of the code change and they want other reviewers to look at it when reasoning about how to vote. Either deciding they want at least one more approval:

\begin{quote}
\emph{``Mmm, the change is so small that you could set +2 here. Just to say that is it OK. But in this case I still think that I would like to set +1. Because I would like, when you set +1, you are saying `looks good to me, but someone else must approve'.''} \textemdash P05
\end{quote}

\noindent \ldots or that their own review is sufficient:

\begin{quote}
\emph{``This is such a small thing so then I feel like this, are two [pairs of] eyes required for this? Except mine? No, here I somehow trust Person D [commit author] and myself so then I think that, yes, we might as well set +2.''} \textemdash P05
\end{quote}

\noindent \ldots or when they are the second reviewer and give the final approval:

\begin{quote}
\emph{``Yes, yes, and here they have voted code review +1, so then I could set +2 actually. Because there is no reason [not to] since I think it looks reasonable.''} \textemdash P09
\end{quote}

\subsubsection{Theme: Understanding Implementation}

This theme includes questions about, i.e., the execution flow through the code, call signatures, variable declarations, comparing the code before and after the change, etc. One participant traces the execution of the code and decides to check it out in their IDE. 

\begin{quote}
\emph{``Then he does a 'findOne'. Then I think I want to look at it in my IDE here. Because I wonder what is going one. Let's see, let's see\ldots''} \textemdash P06
\end{quote}

\noindent There are also general questions about how the code works.

\begin{quote}
\emph{``I'm just trying to understand what, what the change does''} \textemdash P01
\end{quote}

\begin{quote}
\emph{``No, I don't know what the hell it does.''} \textemdash P02
\end{quote}

\noindent Another reviewer traces the execution to finally understand how it fits together.

\begin{quote}
\emph{``Umm, this became a bit strange because here you have, here we get a context from the\ldots or, yes, really from the event-pipeline as a whole that calls this method to handle an activity-started event and there we get a context that can contain a timeout or a cancellation. So we need to\ldots and then I get this, but the transaction as a whole has, does not get any context. But we do send the context in\ldots Ah, right! When we look up, when we fetch a build!''} \textemdash P02
\end{quote}

In other cases, something that looks like a serious mistake makes the reviewer ask why the code does not crash.

\begin{quote}
\emph{``So this seems weird. It should actually, it should crash there with a key violation in the database, I think, because you are not meant to be able to register the same activity on the same build attempt more than once.''} \textemdash P02
\end{quote}

\subsection{Transitions Between Themes}
\label{sectionResultsTransitions}

\begin{figure}[!htb]
\centering
\includegraphics[width=0.98\textwidth]{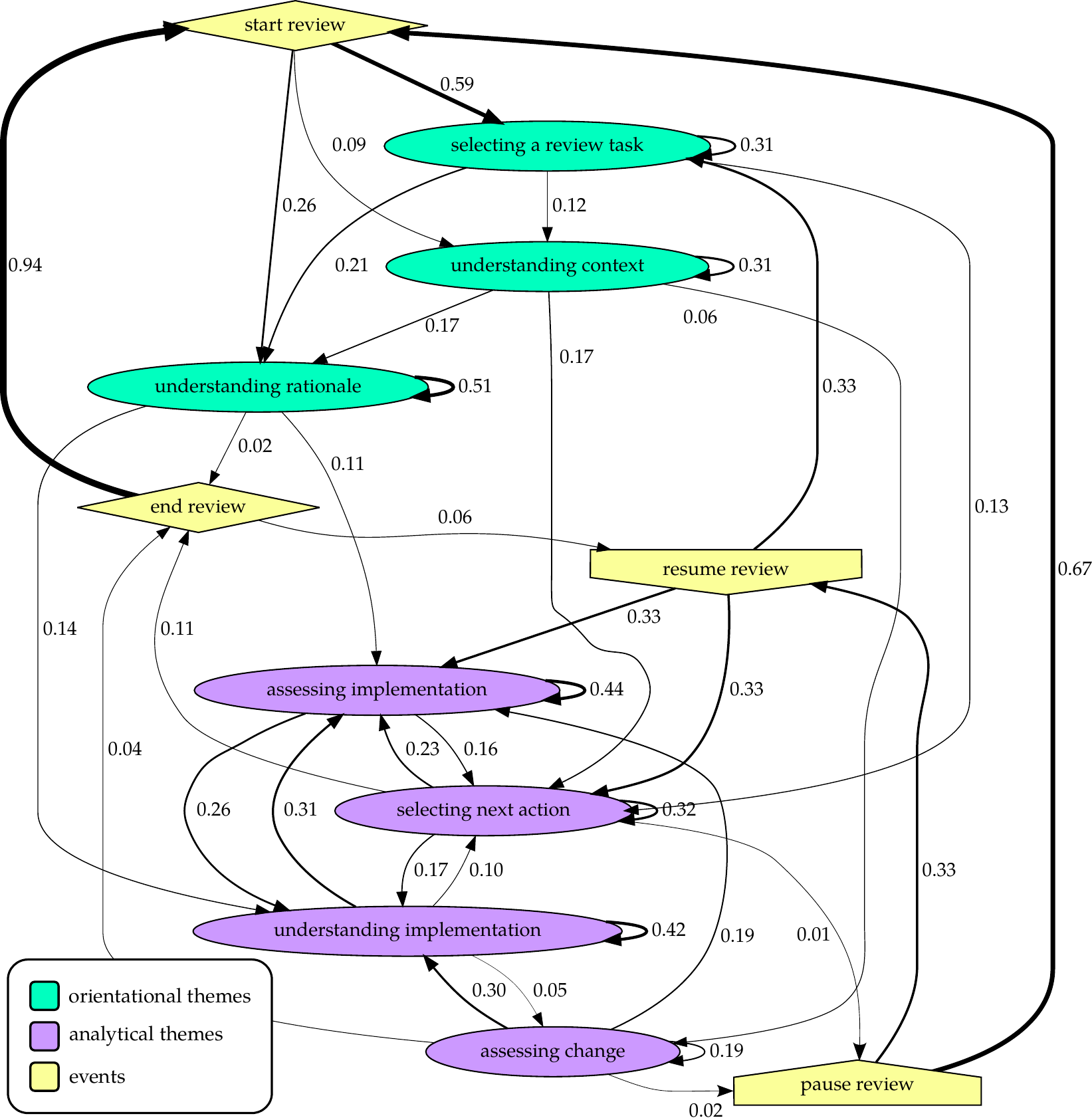}
\caption{Transitions between themes during code review. \textbf{Takeaway:} After starting the review, there is a linear flow through the three orientational themes followed by an iterative loop through the four analytical themes.}
\label{figure:rel_network}
\end{figure}

\begin{figure}[htb]
\centering
\includegraphics[width=0.9\textwidth]{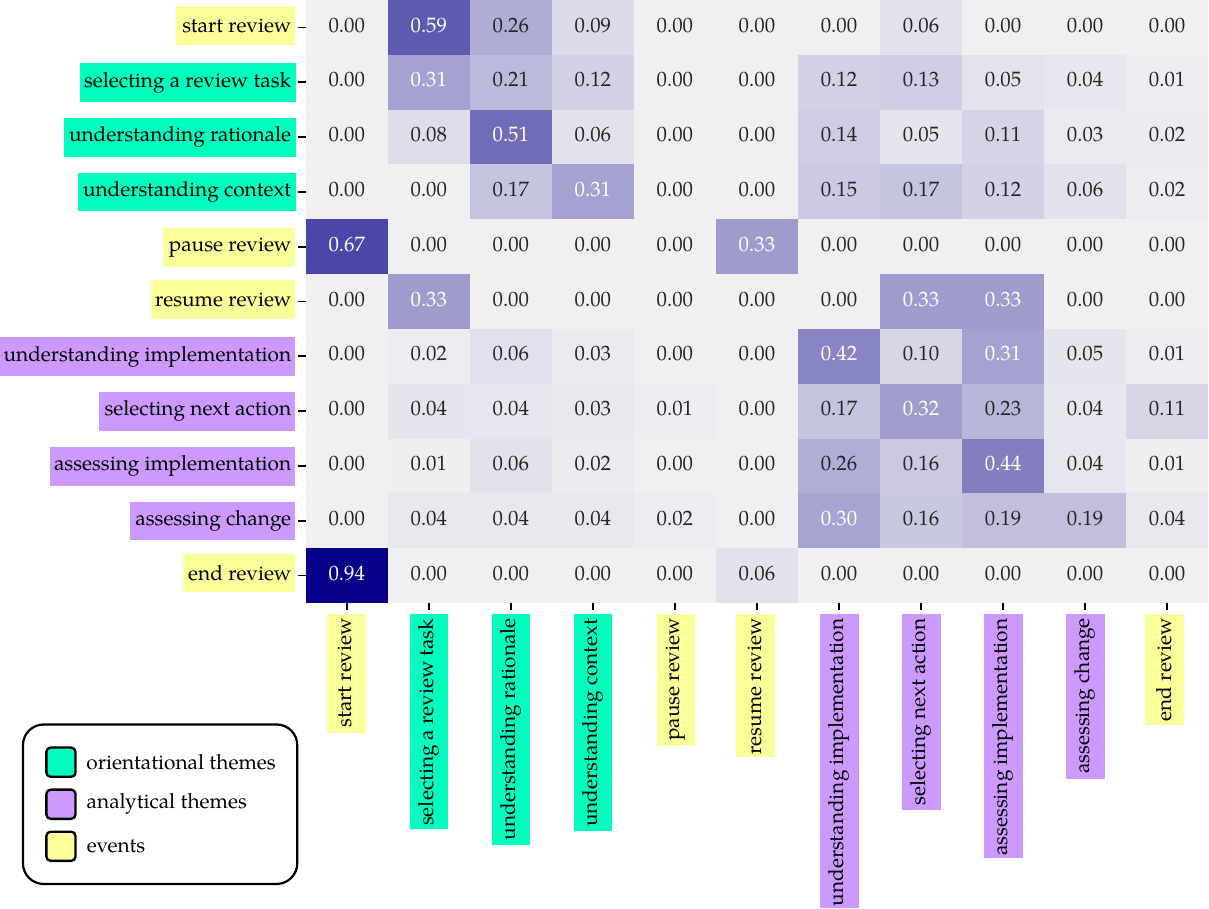}
\caption{Transition probabilities between themes during code review. \textbf{Takeaway:} Three denser clusters appear, one in the top left quadrant between the orientational themes, one in the bottom right quadrant with the analytical themes and one less dense in the top right quadrant showing the transition from orientational to analytical themes. }
\label{figure:rel_transitions}
\end{figure}

In \Figure{figure:rel_network}, we visualize the results of the sequential analysis (\Section{method:data_analysis}) as a sequence diagram. The graph shows the themes and the transition rate of the questions being asked shifting from one theme to another. In addition to the themes, the events `start review', `pause review', `resume review', and `end review' are included as reference points for the process. The 3 most frequent outgoing and incoming edges are plotted. A full list of transition rates can be found in \Figure{figure:rel_transitions}. The transition rates are normalized to 1 over the outgoing edges, while the sum over the incoming edges can be higher or lower. For example, for less frequent themes such as \emph{end review}, which occurs at most once for every review, the sum of the incoming transition rates is much lower than 1.

In the diagram, we can see that, right after starting a code review session, the most common action (60\% of cases) is to select which code change to review. When the review task is selected and its scope and expectations are known, the questions first shift to understanding the context and then to understanding the rationale for the change. If the reviewer has already decided what to review before starting the session, they usually go straight to trying to understand the rationale, maybe because in these cases they already know the context. 

These first three themes are all more orientational in nature, asking about, for example, expectations, author, related work, and rationale. Also note that while the reviewer often iterates on every theme, they never move backward once they have started asking questions on the next theme. We believe that this is because there is a causal link between these questions. You need to have selected what to review before it is meaningful or possible to learn the context and rationale of what you are reviewing. Likewise, if you after understanding context and rationale go back and pick another code change to review, it means that you are ending the current review and starting a new review session. You can see in the diagram that this actually happened in our dataset, although very few times, as indicated by the arc going from \emph{understanding rationale} straight to \emph{end review}.

After that, the reviewer moves on to the remaining four themes; \emph{assessing implementation}, \emph{selecting next action}, \emph{understanding implementation}, and \emph{assessing change}. These themes are more analytical compared to the initial three themes, and revolve around understanding, assessment, and planning. In the sequence diagram, they form a series of connected loops, and we can see that the reviewer moves iteratively between them.

Notably, the fact that the \emph{selecting next action} questions are so central in the iterative loop highlights that code review is not a linear and straightforward activity where the reviewer just reads the diffs of the changed files from beginning to end. Instead, the reviewer constantly plans what to do next and will often revisit the same files and lines many times during a review.

\subsubsection*{Interrupted Reviews}

If a review is interrupted for some reason, in two thirds of the cases it is resumed by going directly to \emph{assessing implementation} or \emph{selecting next action} and then into the iterative loop described above. In one third of the cases, we see that even if the reviewer has read some of the code before they were interrupted, they start over from the beginning and read all of the code again to gain a full picture of the code change and be able to reach a decision on how to vote. Also, even if the reviewer goes straight into assessing implementation, they will often start over and read from the first code diff in the change.

\begin{quote}
\emph{``I have, I prefer to try to review the whole change in one [sitting]. Because if I get half-way and then get interrupted and have to go do something else, yes, I have a tendency to get lost a little bit. You have to go through everything, it is faster the second time but you still have to do it.''} \textemdash Participant 7
\end{quote}

\subsection{Distribution of Themes during Code Review}

\begin{figure}[htb]
\centering
\includegraphics[width=0.99\textwidth]{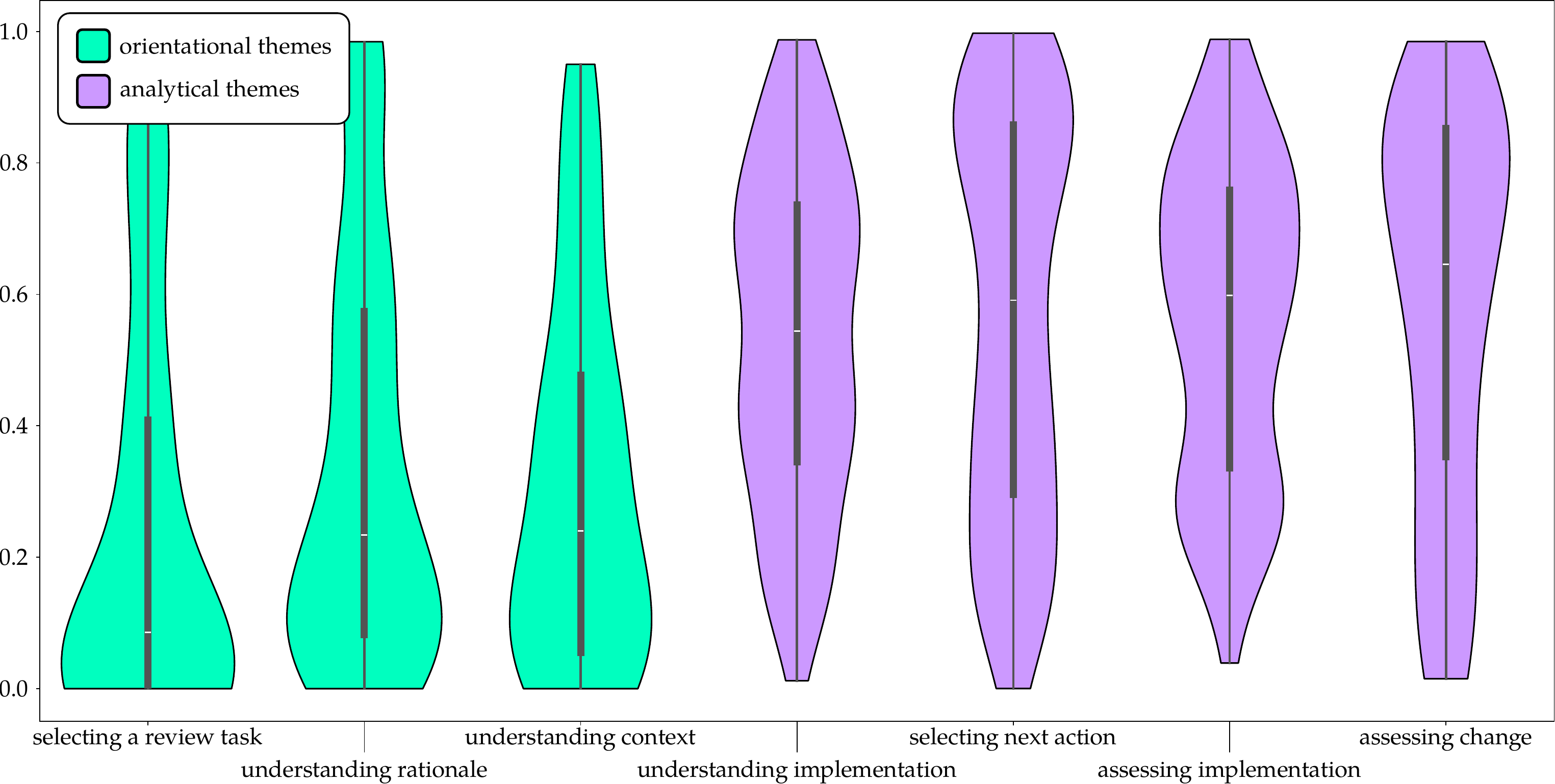}
\caption{Distribution of themes relative to the beginning (0) and end (1) of the code review. \textbf{Takeaway:} The orientational themes and the analytical themes form two groups respectively. Each group having similar timestamp distributions internally, but distinct from the other group.}
\label{figure:category_violins}
\end{figure}

In \Figure{figure:category_violins} we show the distribution of themes over the normalized timeline of a code review where 0 and 1 represent the start and the end of the review session, respectively (see \Section{method:data_analysis}). The themes are sorted ascending by the mean timestamp. 

Supporting the patterns seen in \Section{sectionResultsTransitions}, we see that the themes of the two topics have similar characteristics within each topic but separate characteristics between topics. Sorting by mean timestamp separates the themes into two groups aligned with the topics. The orientational themes \emph{selecting a review task}, \emph{understanding rationale}, and \emph{understanding context} occur from the beginning to about the midpoint of the review with a mean timestamp of $0.2-0.3$. The analytical themes \emph{understanding implementation}, \emph{selecting next action}, \emph{assessing implementation}, and \emph{assessing change} have a timestamp distribution that is centered around the middle and extends to the end of the code review with a mean timestamp of $0.5-0.6$. 

\section{Theory}
\label{secTheory}

Here, we present a theoretical model of code review as a decision-making process, first identifying observed phases and then mapping our observations to the RPD model described by \citet{klein_sources_1998} (\Section{secRPD}).

\subsection{The Two Phases of Code Review}

When thematic, temporal, and sequential analyses of the questions asked during code review are combined, two distinct phases emerge. A linear phase at the beginning of the review that we call \emph{orientation phase}, and an iterative phase from the middle to the end of the review that we call \emph{analytical phase}.

\subsubsection{The Orientation Phase}

\begin{figure}[H]
\includegraphics[width=0.99\textwidth]{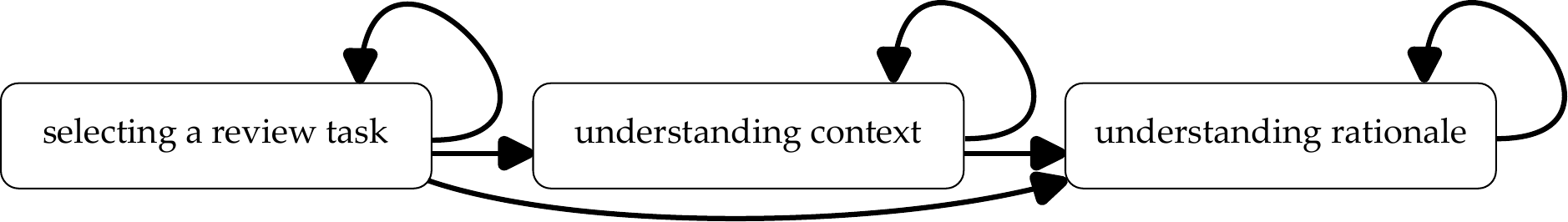}
\caption{Orientation phase of code review. \textbf{Takeaway:} In this phase, reviewers explore each theme with several questions before moving forward to the next. All transitions move forward and the questions are related to rationale, context, and expectations.}
\label{figure:orientation-phase}
\end{figure}

In the orientation phase of code review (see \Figure{figure:orientation-phase}) the reviewer asks about and establishes context, scope, rationale, and expectations. The phase consists of the three themes \emph{selecting a review task}, \emph{understanding context}, and \emph{understanding rationale}. The themes fall under the functional topic of orientational themes, share early mean timestamps in the temporal analysis, and mostly forward transitions in the sequential analysis.

The reviewer begins by asking one or more questions under the \emph{selecting a review task} theme, for example, asking about how big the code change is, which code base it is in, continuous integration (CI) status, priority, and urgency. Next, they continue with the theme \emph{understanding context} asking questions about the code change author, repository, programming language, and type of change (bug fix, feature, refactoring, etc.). Finally, the reviewers ask questions on the theme \emph{understanding rationale}. For example, about the commit message content, issue description, feature requirements, and recent team conversations. Sometimes, the context is well known, from the team stand-up meeting or other recent discussions, and the reviewer skips directly to \emph{understanding rationale}. Each theme can be repeated multiple times with several questions exploring the theme, but once the questions transition to the next theme, the reviewer rarely goes back again.

\subsubsection{The Analytical Phase}

\begin{figure}[H]
\includegraphics[width=0.99\textwidth]{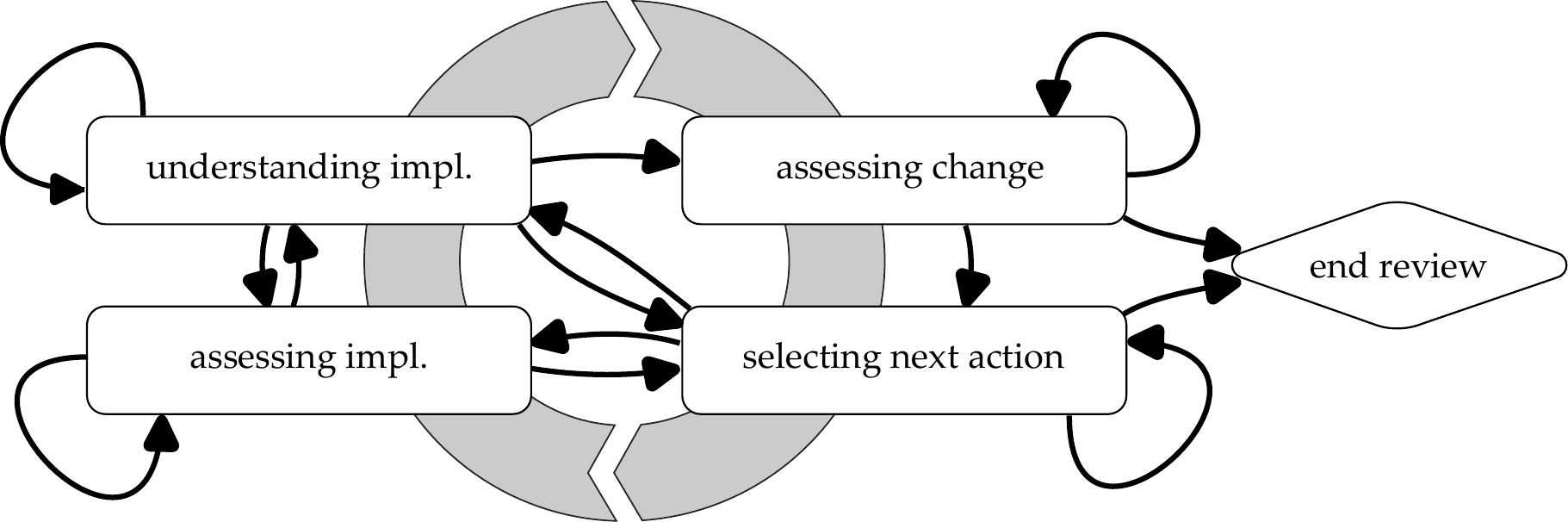}
\caption{Analytical phase of code review. \textbf{Takeaway:} The analytical phase iterates over understanding implementation, assessing implementation, assessing the change, and planning the next step. Both each theme and also the full cycle are repeated several times before the review is done.}
\label{figAnalyticalPhase}
\end{figure}

In the analytical phase of code review (see \Figure{figAnalyticalPhase}), the reviewer iteratively builds an understanding of the changed code, evaluates the implementation and the change as a whole based on this new understanding, and plans what action to take next. The phase consists of the four themes \emph{understanding implementation}, \emph{assessing implementation}, \emph{assessing change}, and \emph{selecting next action}. The themes all fall under the functional topic of analytical themes, share late mean timestamps in the temporal analysis, have cyclic transitions between them in the sequential analysis, and have very few transitions going back to the themes in the orientation phase.

The reviewer could enter the phase on any of the themes. After entering the analytical phase, the most common scenario is that the reviewer iterates through all the themes several times before reaching a \emph{selecting next action} question in which the reviewer decides to end the review. The final question is often about a summary of the comments the reviewer has written or about how to vote regarding the code change.

\subsection{Introducing a Model of Code Review as Decision-Making}
\label{theoryCRDM}

\begin{figure}
    \centering
    \includegraphics[width=0.99\linewidth]{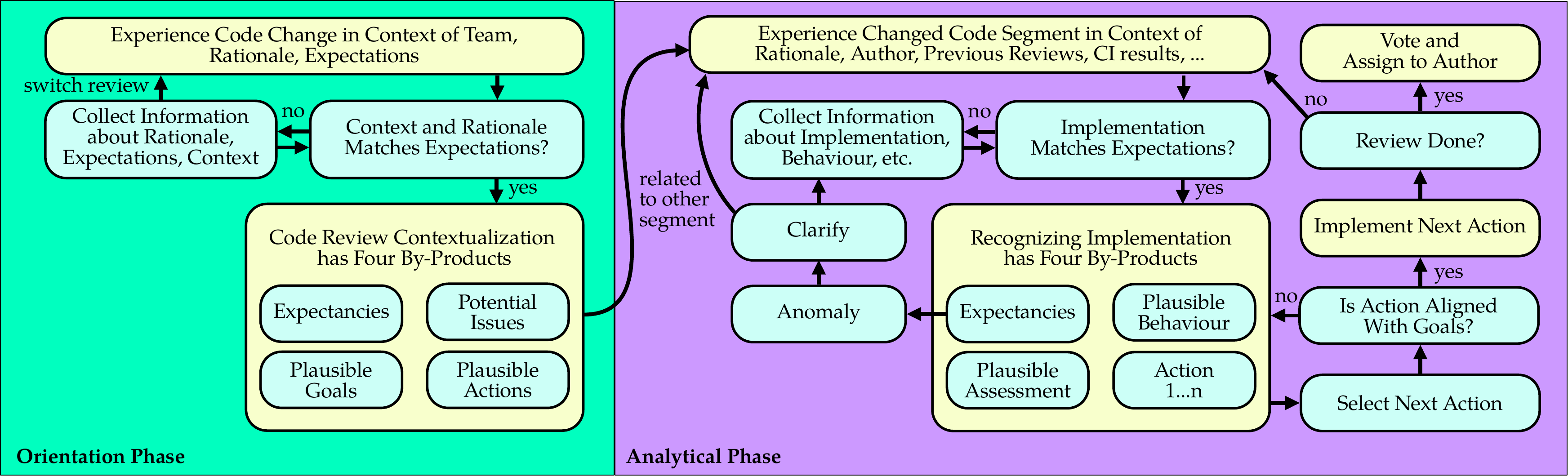}
    \caption{The Code Review as Decision-Making (\textbf{CRDM}) cognitive model. \textbf{Takeaway:} Code review can be modeled as two linked decision-making processes. The first is preparatory and establish context, plausible goals, and plausible actions. The second takes decisions on what actions to take during the review, and how to vote regarding the integration of the code change. It bases these decisions in understanding the change through both implementation details and as part of a larger system.}
    \label{figCRDM}
\end{figure}

The expected outcome of code review is to reach a decision on voting for or against merging the changed code~\citep{sadowski_modern_2018}. On the way there, the participants in the study take several smaller decisions around writing review comments, phrasing of comments, reading external sources, reading changed source code, checking out the code locally, choosing next steps, etc. We can also see that while the cognitive model of code review comprehension~\citep{goncalves_code_2025} is confirmed by our data and can be mapped to the theme \emph{understanding implementation} as a whole and to the mental modeling parts of the theme \emph{selecting next action}, it is just a part of a larger cognitive process.

This leads us to re-frame code review as a kind of decision-making process. Specifically, the orientation and analytical phases in code review can be mapped onto the RPD model introduced by \citet{klein_sources_1998} (\Section{secRPD}). In \Figure{figCRDM}, we illustrate the Code Review as Decision-Making (\textbf{CRDM}) cognitive model mapping the orientation and analytical phases of code review onto one preparatory and one complete RPD model, respectively. Connecting the themes, topics, and phases from our data analysis to the RPD model allows us to construct a cognitive model for the code review task as a whole. A model in which the code reviewers' experience helps them spot potential issues, find confusing or problematic sections of code, formulate effective review comments, look for more information, and vote appropriately for accepting or rejecting the changed code based on their mental simulation of software and team behavior. 

In the orientation phase, the reviewer asks questions about the rationale and context of the changed code in order to build a story around the change and what it tries to do and why. This creates expectancies around the changed code, potential issues to look out for, plausible goals of the review, and plausible actions. If the context or the rationale does not seem to match expectations, the reviewer collects more information, for example, by reading in the issue tracker or asking over the team chat. Since the reviewer usually does not vote or comment during this phase, it is modeled with the preparatory first half of the RPD model.

After the orientation phase, the reviewer enters the analytical phase that can be modeled with a complete cycle through the RPD model. The reviewer experiences the changed code in view of the context established in the orientation phase, reacts to how well it matches their expectations, evaluates if it is in line with the rationale as they have understood it, and chooses their next action. If the changed code triggers a cue for potential issues or does not seem to match the rationale, the reviewer takes action such as writing a review comment, reading external sources, asking for clarification, or voting for rejecting the changed code. When the rationale and the code are coherent, they might perform mental simulation of both what could happen if the code was deployed and of how their colleagues will respond to review comments and voting choices. Finally, after as many iterations as needed, the reviewer implements their decisions on how to vote, which review comments to write, what to communicate in other channels, and what to do next.

Together, the two phases in the CRDM model form a cohesive cognitive process that covers the entire code review task. The orientation phase equips the reviewer with context and rationale for the code change, while the analytical phase enables iterative evaluation, understanding, and action planning. In total, it models code review as a dynamic, experience-driven decision process that includes but goes beyond comprehension.
\section{Discussion}
\label{secDiscussion}

Our results and theoretical modeling show that code review has much in common with decision-making processes, specifically the RPD model by \citet{klein_sources_1998}. Mapping the cognitive phases found during code review onto the RPD model gives us a novel cognitive model (\RQ{1}) of Code Review as Decision-Making; the CRDM model (\Section{theoryCRDM}).
This model can explain and predict some observations from empirical studies of code review. For example, since recognition-primed decision-making requires extensive experience of analogous situations, it can explain how even experienced programmers can take up to a year to become effective at code review in a new workplace~\citep{bosu_characteristics_2015}. It is not just about code comprehension (a well-developed skill for an experienced programmer), but also about building up a mental index of patterns in a new code base and organization. Furthermore, misalignments between current code review tools and developer needs~\citep{soderberg_understanding_2022} could be extrapolated from current code review tools, which center the code diff-view and thereby the \emph{understanding implementation} part of the code review. This design leaves it up to the developer to plan the review, gather decision-making information, explore the context, and understand the rationale, thus showing that current tools are not aligned with all the needs and goals of users. 

The CRDM cognitive model also allows insight into how code review compares to more well-studied processes such as decision-making, reflection, and learning. Being able to relate research results from other disciplines through the theoretical model can inspire future research, give insight into the challenges of code review, and indicate directions for improvements in code review tools (\Section{secFutureWork}). 

\subsection{Insights from Thematic Analysis}

Looking at the levels of topics and themes (\RQ{2}) gives us insight into the relative frequency of questions asked during code review. On topic level, the balance is roughly 70/30 between questions with analytical themes and orientational themes; emphasizing the analytical nature of code review while still pointing out context and orientation as an imprescriptible part.

At the theme level, questions seeking to understand the implementation, rationale, and context account for slightly less than half of the total questions asked. These themes can be related to processes of comprehension, as described by both \citet{letovskyCognitiveProcessesProgram1987} and \citet{goncalves_code_2025}. Also, on this level, these results show that while comprehension is a very important part of code review, there is more involved to complete the task. Planning, decision-making, and assessing account for the other half of the questions asked. Notably, a significant part of the analytical work during code review consists of \emph{selecting next action}, which is concerned with planning and decision-making and in a way is a metacognitive theme.

\subsection{Classification of Questions During the Code Review}

When analyzing topics and themes as a sequential process during code review (\RQ{3}), we see interesting patterns in how participants move between themes in their questions. \citet{letovskyCognitiveProcessesProgram1987} classifies questions asked during a cognitive process into five groups; \emph{Why}, \emph{How}, \emph{What}, \emph{Whether}, \emph{Discrepancy}. From the thematic and sequential analysis, we see that the first orientation phase is dominated by the \emph{Why} and \emph{What} questions exploring the rationale and context of the code review. In the analytical phase, this emphasis shifts to \emph{How}, \emph{Whether}, and \emph{Discrepancy} questions that explore how the code works, whether it could have desired or undesired behaviors, and assessing it for discrepancies with expectations and code guidelines. That the questions are different in the two phases we conjecture, also when applying Letovsky's classification system in addition to our own thematic coding, supports the division into orientation and analytical phases.

\subsection{Comparison to Previous Models}

Existing models for code review focus on and describe the organizational process~\citep{sadowski_modern_2018,bacchelli_expectations_2013,mcintosh_empirical_2016}, but to our knowledge only one model of the cognitive process has been presented; the code review comprehension model of \citet{goncalves_code_2025}, which deliberately only models a part of the entire code review process; comprehending code changes. For this part of the review, our results support their findings. Especially, the reviewer iteratively uses information sources, a knowledge base, and their own mental model to drive comprehension during the code review. The CRDM model then expands the scope significantly by describing orientation processes before starting to read the changed code and the recurring decision-making needs throughout the review. 

\subsection{Future Work}
\label{secFutureWork}

\begin{figure}[htb]
\centering
\includegraphics[width=0.58\linewidth]{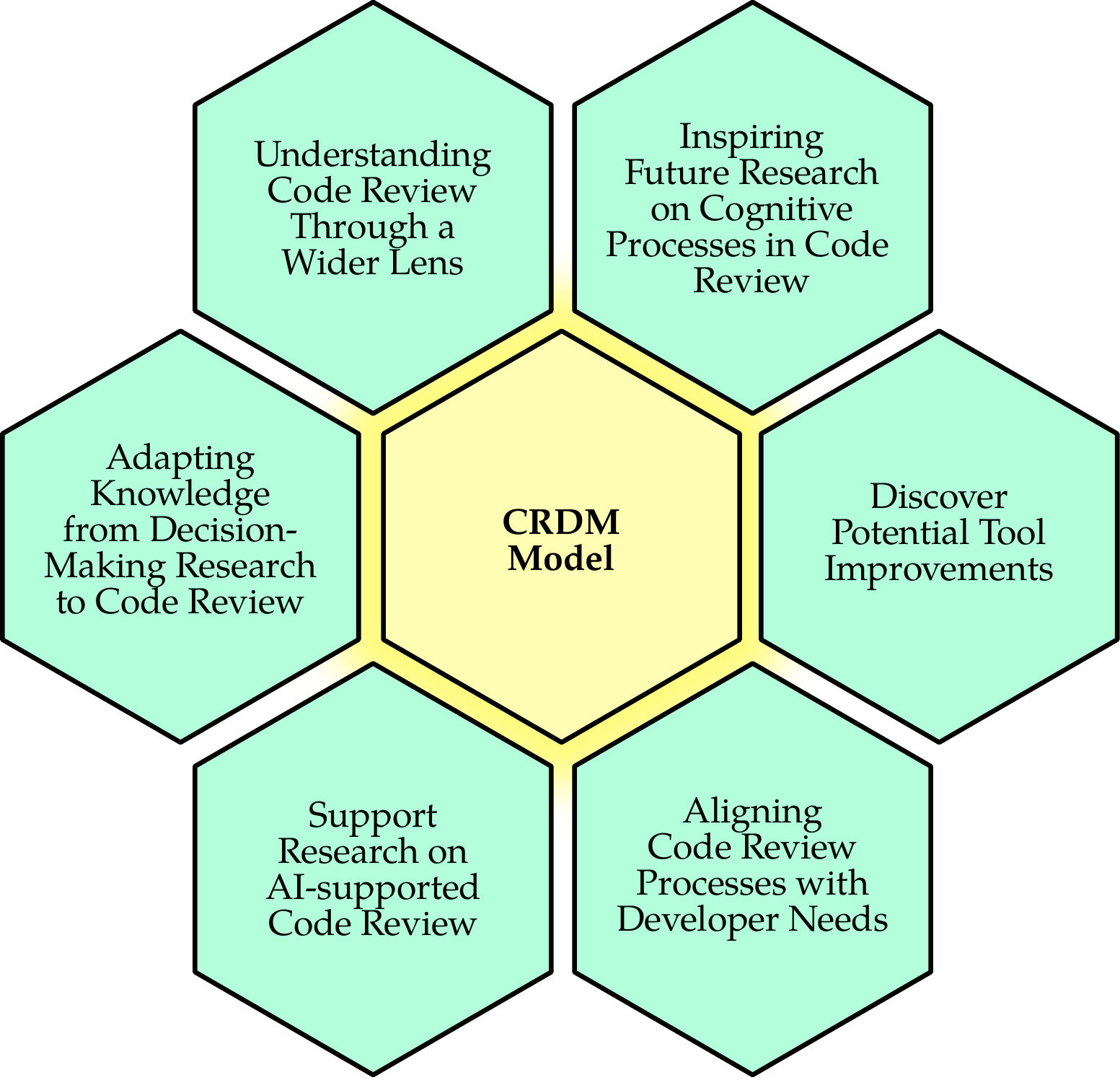}
\caption{Illustration of potential applications of the CRDM cognitive model. Each surrounding hexagon represents a potential application of the model, such as adapting knowledge from decision-making research, improving code review tools, and supporting future research. The center hexagon represents the CRDM cognitive model itself.}
\label{figHexagons}
\end{figure}

As illustrated in \Figure{figHexagons}, the CRDM model and the results leading up to the model can have several applications for future research and development.

\paragraph{Adapting Knowledge from Decision-Making Research}

Given the similarities between code review and decision-making processes, code review tools could learn a lot from decision support systems (DSS). \citet{liuIntegrationDecisionSupport2010} presents a meta-study of over 100 papers on DSS and shows the advantages of Integrated Decision Support Systems (IDSS) where the decision support is integrated into existing systems and processes. This would be an interesting way forward for code review tools. For example, knowledge graph-based or agentic IDSS could be integrated into existing tools to support developers throughout their workflow and make code review more effective and efficient.

\paragraph{Discover Potential Tool Improvements}

Today's code review tools come from a legacy of code inspection meetings and have, on a feature level, not changed much since the first ever software for code review, ICICLE, was introduced in 1990~\citep{brothers_icicle_1990}. By analyzing the needs of developers during the different steps in the CRDM model, the features of popular code review tools~\citep{githubFeat2025, gerritUserGuide2025}, and recent research on misalignment between code review tools and user goals~\citep{soderberg_understanding_2022} we conclude that today's code review tools have significant room for improvement.

Reevaluating code review tools from the lens the CRDM model, or incorporating ideas and features from DSS software discussed above could give benefits to development teams around the world. In particular, the tools' focus on the code diff view and inline review comments centers its support mostly around the two themes \emph{understanding implementation} and \emph{assessing implementation}. Overall, the \emph{orientation phase} as a whole receives less support, and users often choose to leave the code review tool and look for information in issue tracker, team chat, external documentation, etc., to better understand the context and rationale. 

\paragraph{Support Research on AI-supported Code Review}

As discussed in \Section{secIntroduction}, more and more research and development in academia and industry is directed towards supporting or replacing code review with AI models or agents. Many researchers are calling out the importance of preserving the human perspective and emphasizing AI technology that supports rather than replaces current software engineering practices \citep{russo_generative_2024,heander_support_2025}. A cognitive model can be a useful foundation in finding areas where current tools give insufficient support and where AI models and agents could augment the capabilities of the human engineers. Training the CRDM model into an agentic workflow could also provide pre-emptive support and information and guide developers through the code review task.

\paragraph{Understanding Code Review Through a Wider Lens}

In much of the literature today, code review is seen as a process of comprehension and defect finding. While that is certainly part of the truth, we would like to challenge and expand this view and treat code review as more closely related to decision-making. By applying this wider lens, we believe that there are new insights and perspectives to be found that might, for example, change the way we teach code review to new developers and the way we design tools.

\paragraph{Aligning Code Review Processes with Developer Needs}

As well as misalignments between the developer needs and the tools used, \citet{soderberg_understanding_2022} also finds misalignments between the process itself and responsibilites and outcomes of code review. Perhaps there are ways to adapt the accepted code review processes to better match the needs and cognitive process of the developers involved in the process.

\paragraph{Inspiring Future Research on Cognitive Processes}

To our knowledge, few cognitive models of code review have been published today. We hope by presenting methodology and results from building the CRDM model that we can inspire future research in studying software engineering processes from a cognitive lens. Certainly, there are parts of general code review, such as which actions reviewers take to resolve their questions, that warrant further study. It could also be illuminating to study code review in different kinds of companies, open source projects, development methodologies, and team sizes.

\section{Threats to Validity}

LeCompte and Goetz present a comprehensive investigation of both internal and external threats to be considered for ethnographic research~\citep{lecompte_problems_1982}. They separate threats to reliability, which concerns to what degree the study is reproducible, and to validity, describing the accuracy of the conclusions in relation to empirical reality. In the following subsections, we use the threats identified by LeCompte and Goetz to analyze the threats in this study. 

\subsection{External Reliability}

The external reliability of an ethnographic study is affected by \emph{position of the researcher in the study}, \emph{choices of the informants}, \emph{social situations and conditions}, \emph{analytic constructs and premises}, and \emph{methods of data collection and analysis}~\citep{lecompte_problems_1982}.
The researcher conducting the field work has extensive experience in software engineering and code reviews, contributing to good rapport between the researcher and the participants. This experience facilitated open sharing of thoughts and experiences as well as understanding of specific terms, jargon, and practices (\Section{sectionMethod}). In the research team, we also have one more member with long industry and code review experience, as well as one researcher without industry experience but with extensive knowledge in human factors, cognition, and interaction studies. This contributed to building an understanding and theory that is valid from both an insider and an outsider perspective.

Regarding informant choices, there is inherent bias in the fact that people who volunteer to participate in ethnographic research studies are introspective and insightful about their own thinking and actions to a greater degree than the average in most groups~\citep{lecompte_problems_1982}. In our case, choosing to work with the outward-facing software tools department that collaborates with external open source communities and with all other internal departments gives a bias towards people who are communicative, outgoing, and used to describing their work to outsiders. This increased the depth and detail we could achieve in our collected data. We think the risk that less extroverted participants would follow a fundamentally different code review process is small.

All participants and teams in this study have strong similarities (\Section{sectionMethodParticipants}). Developers with, for example, a different cultural and educational background working in large teams using waterfall methodology might have a different approach to code reviews. We have tried to mitigate this risk by actively choosing participants ranging from inexperienced to very experienced and with different roles in their teams. 

The social context and setting for the code reviews were at the developers' regular workplace, on their own computer, monitor, and desk to create conditions for realistic results.
%
Our analytical constructs are based on our constructivist epistemological view, as well as the process codes used to encode the transcripts (\Section{sectionMethod}). The process codes follow accepted coding practices and are published in the replication package, see \Section{matterReplicationPackage}
Finally, data collection and analysis were performed using common practices in qualitative studies, such as audio recording, transcription, process coding, and statistical analysis~\citep{Charmaz14, saldana_2015}. Transcribing the recorded interviews carry the risk of subtly shifting the meaning, since spoken and written language is interpreted slightly differently and the transcriptions lack prosody and tone of voice. We mitigated this by adding notes in the transcription where the meaning would otherwise be ambiguous.

\subsection{Internal Reliability}

The internal reliability of an ethnographic study is affected by \emph{low-inference descriptors}, \emph{multiple researchers}, \emph{participant researchers}, \emph{peer examination}, and \emph{mechanically recorded data}~\citep{lecompte_problems_1982}.
To achieve low-inference descriptors, the source materials for analysis in the study were verbatim transcriptions of the recorded code review sessions with little to no inference.
Multiple researchers were involved in the interpretation of the data. 
Two researchers independently did the thematic coding of the material. All three researchers discussed process coding, thematic coding, and topics until agreement~\citep{saldana_2015}.
The process coding and its interpretation were further verified with the participants through a member-checking workshop (\Section{methodMemberChecking}).
For peer examination, we note that \citet{goncalves_code_2025} describes a process, albeit within a more narrow scope, that confirms our model in the parts where they overlap.
Finally, data were recorded using digital dictaphones for voice clarity, and the original recordings are archived at the university.

\subsection{External Validity}

The external validity of an ethnographic study is affected by \emph{selection effects}, \emph{setting effects}, \emph{history effects}, and \emph{construct effects}~\citep{lecompte_problems_1982}. 
To address selection effects, we selected participants with different age, experience, and roles, but working for the same company and team. They follow the same or very similar code review guidelines. This contributes to results from different participants being comparable.
Since the observations were carried out in the same office with members of the same development team, the settings were very similar. We think that the social effects of the setting, group, and researcher are comparable between the interviews.
Regarding history effects and construct effects, all participants have a comparable cultural and educational background, which contributes to the validity of comparing their data. 
Further, the same process-coding and thematic coding was used for all recordings and participants, again contributing to comparability between participants.

\subsection{Internal Validity}

The internal validity of an ethnographic study is affected by \emph{history and maturation}, \emph{observer effects}, \emph{selection and regression}, \emph{mortality}, and \emph{spurious conclusions}~\citep{lecompte_problems_1982}.
In relation to history and maturation, code review was a well established and mature practice in the development team that participated in the study, and their guidelines remained the same throughout the field work.
We may have had observer effects in that participants might have put in more effort than usual into the code reviews. That is, participants may have spent more time in code review and may have been more meticulous with comments and approvals, to be perceived as competent by the researcher and their peers. During field work, we tried to mitigate observer effects by being neutral and curious about any approach the participants took.
While we did select as diverse participants as possible from the members of the participating development teams, they do have similar cultural and educational backgrounds. It cannot be ruled out that participants with different culture, education, role, employer, role, etc. would also have different strategies during code review.
No participants left the study (or the team) during the field work.
Finally, conclusions and theories were built bottom-up from the results of thematic and statistical analysis to avoid drawing spurious conclusions from our field observations.

\section{Conclusions}
\label{sec:conclusions}

We studied questions asked during code review using an ethnographic think-aloud study combined with interviews (\Section{sectionMethod}). The study included 10 participants and a total of 34 code reviews. We performed thematic analysis of the transcribed interviews followed by temporal and sequential analysis. Through this analysis, we discovered patterns in the kinds of questions that reviewers asked during code reviews; when the questions were asked, and how the questions connected to each other.

From our thematic analysis we identified 2 topics containing a total of 7 themes (\Figure{fig:theme_map}). Temporal and sequential analysis indicates that code review can be modeled by two phases (\Section{secTheory}), a linear \emph{orientation phase} (\Figure{figure:orientation-phase}) followed by an iterative \emph{analytical phase} (\Figure{figAnalyticalPhase}). During the orientation phase, the reviewer seeks information about the expectations, rationale, and context of the code change within and outside the code review tool. Once those factors are understood, the reviewer enters the assessment phase. Here, they iterate by seeking to understand the implementation, assessing the change, assessing the implementation, and planning their next action until the code review is finished.

The similarities in the dynamics during these two phases with decision-making processes in general, and in particular the RPD model defined by \citet{klein_sources_1998}, lead us to propose the Code Review as Decision-Making (CRDM) cognitive model (\Section{theoryCRDM}). In this model, we reframe code review as a decision-making process, providing new perspectives on the practice, its effects, avenues for future research, and ideas on how tools can evolve to support code review in a better way.

\backmatter

\section{Notes}

\bmhead{Acknowledgments}
The authors would like to thank the following funders who partly funded this work: the Swedish strategic research environment ELLIIT, the Swedish Foundation for Strategic Research (grant nbr. FFL18-0231), the Swedish Research Council (grant nbr. 2019-05658), and the Wallenberg AI, Autonomous Systems and Software Program (WASP) funded by the Knut and Alice Wallenberg Foundation. The authors also thank the study participants for their enthusiastic collaboration and open sharing of their thoughts and knowledge.

\bmhead{Author contributions}
All authors contributed to the conception and design of the study. The first author performed material preparation, data collection, transcription, data analysis, data visualization, and illustration. All authors contributed to the thematic analysis of the transcribed data. The first draft of the manuscript was written by the first author and all authors commented on previous versions of the manuscript. All authors read and approved the final manuscript.

\bmhead{Data availability}
Anonymized data on the level of process coding that support the findings of this study are openly available in the replication package below. Due to sensitivity reasons, full recordings and transcriptions are not openly available and are available from the corresponding author upon reasonable request.

\bmhead{Code availability}
\label{matterReplicationPackage}
The program code for data analysis is available in the replication package at: 
\href{https://doi.org/10.5281/zenodo.15758266}{https://doi.org/10.5281/zenodo.15758266}

\bibliography{refs}


\end{document}